\documentclass{article}
\usepackage[super,compress]{cite}

\usepackage{epsfig}
\usepackage{xcolor}
\usepackage{cancel}

\def\bv{\mbox{\bf v}}

\def\bphi{\mbox{\boldmath $\phi$}}

\def\balpha{\mbox{\boldmath $\alpha$}}
\def\bbeta{\mbox{\boldmath $\beta$}}

\def\bsigma{\mbox{\boldmath $\sigma$}}

\def\bG{\mbox{\bf G}}

\def\bPhi{\mbox{\boldmath $\Phi$}}

 \def\tfrac#1#2{{\textstyle{{#1}\over{#2}}}}

\def\half{\tfrac{1}{2}}
\def\third{\tfrac{1}{3}}

\def\sixth{\tfrac{1}{6}}

\def\phivac{\bPhi_{\mathrm{\scriptscriptstyle{VAC}}}}

\def\Tr{\mathop{\rm Tr}\nolimits}

\usepackage{amssymb}

\begin{document}
\begin{titlepage}

\begin{center}
{\Large {\bf A vacuum transition in the FSM with a possible new take on 
the horizon problem in cosmology}}
\end{center}

\begin{center}

\vspace{.5cm}

{\large Jos\'e BORDES \footnote{Work supported in part by Spanish MINECO under grant PID2020-113334GB-I00
and PROMETEO 2019-113 (Generalitat Valenciana).}}\\
jose.m.bordes\,@\,uv.es \\
{\it Departament Fisica Teorica and IFIC, Centro Mixto CSIC, Universitat de 
Valencia, Calle Dr. Moliner 50, E-46100 Burjassot (Valencia), 
Spain}\\
\vspace{.2cm}
{\large CHAN Hong-Mo}\\
hong-mo.chan\,@\,stfc.ac.uk \\
{\it Rutherford Appleton Laboratory,\\
  Chilton, Didcot, Oxon, OX11 0QX, United Kingdom}\\
\vspace{.2cm}
{\large TSOU Sheung Tsun}\\
tsou\,@\,maths.ox.ac.uk\\
{\it Mathematical Institute, University of Oxford,\\
Radcliffe Observatory Quarter, Woodstock Road, \\
Oxford, OX2 6GG, United Kingdom}

\end{center}

\vspace{.3cm}

\begin{abstract}

The framed standard model (FSM), constructed to explain the empirical mass 
and mixing patterns (including CP phases) of quarks and leptons, in which 
it has done quite well \cite{tfsm,cpslept,cpslash}, gives otherwise the 
same result \cite{higgcoup} as the standard model (SM) in almost all areas 
in particle physics where the SM has been successfully applied, except for 
a few specified deviations such as the $W$ mass \cite{zmixed} and the 
$g - 2$ of muons \cite{fsmanom}, that is, just where experiment is showing 
departures from what SM predicts.  It predicts further the existence of 
a hidden sector \cite{cfsm} of particles some of which may function as 
dark matter.  In this paper, we first note that the above results involve, 
surprisingly, the FSM undergoing a vacuum transition (VTR1) at a scale 
of around 17 MeV, where the vev's of the colour framons (framed vectors 
promoted into fields) which are all nonzero above that scale acquire some 
vanishing components below it.  This implies that the metric pertaining 
to these vanishing components would vanish also.  Important consequences 
should then ensue, but these occur mostly in the unknown hidden sector 
where empirical confirmation is hard at present to come by, but they give 
small reflections in the standard sector, some of which may have already 
been seen.  However, one notes that if, going off at a tangent, one imagines 
colour to be embedded, Kaluza-Klein fashion, into a higher dimensional 
space-time, then this VTR1 would cause 2 of the compactified dimensions 
to collapse.  This might mean then that when the universe cooled to the 
corresponding temperature of $10^{11}$ K when it was about $10^{-3}$ s 
old, this VTR1 collapse would cause the 3 spatial dimensions of the 
universe to expand to compensate.  The resultant expansion is estimated, 
using FSM parameters previously determined from particle physics, to be 
capable, when extrapolated backwards in time, of bringing the present 
universe back inside the then horizon, solving thus formally the horizon 
problem.  Besides, VTR1 being a global phenomenon in the FSM, it would 
switch on and off automatically and simultaneously over all space, thus 
requiring seemingly no additional strategy for a graceful exit.  However, 
this scenario has not been checked for consistency with other properties 
of the universe and is to be taken thus not as a candidate solution of 
the horizon problem but only as an observation from particle physics 
which might be of interest to cosmologists and experts in the early 
universe.  For particle physicists also, it might serve as an indicator 
for how relevant this VTR1 can be, even if the KK assumption is not made.

\end{abstract}

\end{titlepage}

\newpage
\section{Preamble}

Since the narrative is eventually to bring together two disparate 
and seemingly unrelated topics, it is probably worthwhile to begin 
with a brief introduction to the two main protagonists and an 
outline of the plot.

On one side, then, is the horizon problem, well known in cosmology.  
The evolution of the universe in its later stages seems quite 
well-understood in terms of Einstein's relativity via the Friedmann 
equations \cite{friedmann1,friedmann2} leading to the Hubble expansion
\footnote{There is a huge amount of literature on this topic.  In 
particular, we ourselves have learnt much from the book by Steven 
Weinberg \cite{weinbergcos}, various review articles from the Particle 
Data Group website \cite{pdg}, and from \cite{knobel}.}.  The consequent 
prediction of the cosmic microwave background (CMB) and elucidation 
of the nucleosynthesis process are among the greatest achievements 
of twentieth century physics.  Later experimental results, however, 
have revealed what could be a flaw in this picture.  The cosmic 
microwave background is seen to be isotropically uniform to 
an accuracy of about $10^{-5}$, which means that different parts of 
the present sky must have been in causal contact at an earlier 
stage.  However, if one extrapolates the above picture backwards in 
time, one finds that at the decoupling time when the CMB was formed 
and the universe was about half a million years old, the portion of 
the universe that we now see would be many (about $10^5$) times 
larger than the horizon at that time, meaning that the microwave 
background would have been formed from $10^5$ causally 
disconnected regions.  We have thus an apparent contradiction, called 
in the literature the ``horizon problem''.  This is one of the 
problems, perhaps the main one \cite{weinbergcos}, that the idea of 
inflation\footnote{This is a very active field in cosmology, with a huge literature. Some early 
papers are \cite{inflation1,inflation2a,inflation2b,inflation3}.} was historically suggested to address.  
Whether the inflation idea has succeeded, and if it has, which version 
of its applications has suceeded the most, are still hotly debated by 
experts, among whom unfortunately we cannot count ourselves.

On the other side is a vacuum transition in the FSM (framed standard 
model) of which we ourselves are responsible.  The FSM is a framed 
version of the standard model (SM) of particle physics in which the 
frame vectors in internal symmetry space are promoted into fields.  
It was constructed with the hope of offering an explanation for the 
mass and mixing patterns of quarks and leptons observed in experiment, 
which the SM takes as empirical inputs.  In this, the FSM has done 
rather well, as can be seen later in Table \ref{tfsmfit}. A surprising 
consequence of that explanation however is that the FSM should undergo 
a vacuum transition (VTR1) at a scale of around 17 MeV, in which 2 
components of a particular metric in internal colour space vanish,  
which, if true, would have serious consequences in particle physics. 
However, these are mostly in the hidden sector, which the FSM also 
predicts, but is at present little known.  Empirical confirmations 
of the VTR1 are thus hard to come by.  Nevertheless, one has noted 
two salient experimental facts which may be reflections of VTR1 in 
the standard sector.  As these results are still relatively new and 
little known, a more extended outline will be given in the 2 following 
sections for the reader's convenience.

So far, our two characters, the horizon problem in cosmology and the 
vacuum transition in the FSM, have operated in two different areas 
which do not overlap.  However, if one imagines the FSM scheme to be 
embedded, Kaluza-Klein fashion, into an enlarged multi-dimensional 
space-time, then it can lead to interesting cosmological effects.  
In particular, as the early universe cooled and the temperature fell 
to the VTR1 scale of around 17 MeV (about $2.0 \times 10^{11}$ K when 
the universe was around $2.5 \times 10^{-3}$ s old, long before 
the CMB was formed), it would appear that 2 of the 3 compactified 
dimensions corresponding to colour in the Kaluza-Klein space-time 
would shrink from a ${\rm TeV}^{-1}$ size quite abruptly to zero. 
This would cause, we think, the standard 3-D space to expand by an 
estimated amount capable, when extrapolated backwards in time, of 
bringing the present universe back into the then horizon, thus 
solving perhaps, at least formally, the said horizon problem. 
   
The purpose in this paper is two-fold: first, to make known to the 
community the probable existence of the VTR1 transition which we 
think might be of much physical significance, and secondly, to 
bring to the notice of experts in cosmology the observation in the 
preceding paragraph, for them to consider whether it could be of 
use for constructing a realistic model for the evolution of the 
early universe, a question we would much like to ask but are at 
present incompetent to answer for ourselves.

The paper is organized in such a way that readers who are interested 
mostly in the horizon problem can go directly to section 6 at first 
reading, and return to the intervening sections only if they wish to 
find out how in the FSM these results are derived.

\section{The framed standard model}

The FSM was constructed by framing the SM of particle physics, namely 
by incorporating the frame vectors (or vielbeins) in colour and flavour 
space as dynamic variables (``framons'').  Its initial purpose was to 
understand firstly, why there should be 3 generations of quarks and 
leptons, and secondly, why these should fall into the hierarchical mass 
and mixing patterns seen in experiment, which are taken in the SM  
as empirical inputs.  This the FSM has done fairly well in giving, as 
dual colour, the fermion generations (hence 3 and only 3 generations), 
then next not only a picture of how the peculiar mass and 
mixing patterns could have come about, but even quite a decent fit to 
the existing mass and mixing data of quarks and leptons \cite{tfsm}, as 
will be seen later in Table \ref{tfsmfit}.  

In addition, the FSM has acquired the following surprise bonuses for 
which it was not initially intended:
\begin{itemize}
\item {\bf [CP]} a new unified treatment of CP physics \cite{cpslept,
cpslash}, transforming the theta-angle term of the strong CP problem 
into the Kobayashi-Maskawa (KM) phase in the CKM matrix for quarks, 
thus solving simultaneously both the strong CP problem {\it per se} 
and its ``weak'' counterpart of where the Kobayashi-Maskawa phase in 
the CKM matrix originates.  The same extends also to the lepton sector 
to give the CP-violating phase in the PMNS matrix now being measured 
by neutrino experiments \cite{K2K}.
\item {\bf [SM]} the same result as the standard model over almost all 
the range in which the SM has been applied with success \cite{higgcoup},
\end{itemize} 
except for
\begin{itemize}
\item {\bf [DV]} some deviations and anomalies, such as in the $W$ mass 
\cite{zmixed} and in muonic $g - 2$ \cite{fsmanom}, precisely where 
the latest experiments \cite{Wmass,gminus2} are showing likely departures 
from standard model projections,
\end{itemize}
and
\begin{itemize}
\item {\bf [HS]} the prediction of a hidden sector \cite{cfsm} which 
communicates but little with the standard sector we know and contains 
elements that can play the role of dark matter.
\end{itemize}

We shall not attempt to summarize here what is of necessity a longish 
story how these results come about.  A fairly recent review on this 
material can be found in \cite{variym}, except for the item {\bf [SM]} 
in the above list which has only lately and belatedly been recognized 
\cite{higgcoup}.  We shall just highlight below a few features of the 
FSM scheme needed to make the present discussion intelligible.

Frame vectors can be thought of as the columns of a transformation 
matrix relating a local frame (that is, depending on space-time points) 
to a global reference frame (that is, independent of space-time points).  
Hence, when promoted into (framon) fields, like vierbeins in gravity, these 
will carry 2 sets of indices, one corresponding to the local and one 
to the global frame.  Framons form thus double representations, both 
of the local gauge transformations and of the global transformations of 
the reference frame.  And the resultant framed theory incorporating 
the framons as dynamical variables as the FSM purports to be has then 
to be invariant under local gauge transformations as well as global 
transformation of the reference frames.  The SM has local symmetry:
\begin{equation}
G = u(1) \times su(2) \times su(3)
\label{G}
\end{equation}
The FSM has thus to be invariant under $G \times \tilde{G}$, with global:
\begin{equation}
\tilde{G} 
  = \tilde{u}(1) \times \widetilde{su}(2) \times \widetilde{su}(3).
\label{Gtilde}
\end{equation}
Here, we use a tilde to distinguish the global from the local symmetry, 
although as abstract groups they are identical. 

The framons in FSM are double representations of $G \times \tilde{G}$,
as distinguished from, for example, the gauge fields which are representations 
of only the local symmetry $G$.  Explicitly, the framons in FSM appear 
as:
\begin{itemize}
\item (FF) the flavour (``weak'') framon:
\begin{equation}
\balpha \Phi = \left( \alpha^{\tilde{a}} \phi_r^{\tilde{r}} \right);
    \ \ r = 1, 2; \tilde{r} = \tilde{1}, \tilde{2}; \ \  \tilde{a} = 
    \tilde{1}, \tilde{2}, \tilde{3},
\label{fframon}
\end{equation}
\item (CF) the colour (``strong'') framon:
\begin{equation}
\bbeta \bPhi = \left( \beta^{\tilde{r}} \phi_a^{\tilde{a}}
    \right); \ \  a = 1, 2, 3; \ \  \tilde{a} = \tilde{1},
    \tilde{2}, \tilde{3}; \ \  \tilde{r} = \tilde{1},
    \tilde{2}.
\label{cframon}
\end{equation}
\end{itemize}
\noindent
The factor $\balpha$ is a global (independent of 
space-time point $x$) unit vector in 3-dimensional dual colour 
space which is to play the role of generations for quarks and 
leptons.  The factor $\Phi$ is a $2 \times 2$ matrix 
field (dependent on space-time point $x$) which is, as will be 
explained later, essentially the standard Higgs field.  It is 
the presence of the factor $\balpha$, carried by the flavour 
framon into the mass matrices of quarks and leptons, which will 
lead in the FSM to the hierarchical intricacies of their mass 
and mixing patterns.  Similarly, $\bbeta$ is a global 2-D unit 
vector in dual flavour space, and $\bPhi$ a $3 \times 3$ matrix 
field, the columns of which, labelled by $\tilde{a}$, transform 
as triplet under colour $su(3)$, while its rows transform as 
anti-triplets under dual colour $\widetilde{su}(3)$.  These are 
new ingredients introduced by the FSM with no SM analogues.

That the framon separates into a flavour and a colour part means that 
the FSM has chosen for the framon the sum representation over the 
product representation for the product symmetry $su(2) \times su(3)$, 
that is, explicitly the representation $\bf{1} \times [\bf{2} + \bf{3}]$,
namely: doublet for flavour plus triplet for colour, times a phase 
for $u(1)$.  In principle, in $G$ of (\ref{G}), for the product of each 
pair of the 3 components, the sum or product representation can be 
chosen for the framon.  That $\bf{1} \times [\bf{2} + \bf{3}]$ was 
chosen was because it was the only one that was found to work, but 
it happens also to require the smallest number of independent scalar 
fields to be introduced.  

Further, the flavour framon in {\bf [FF]} above contains 2 flavour 
doublets of scalar fields, but only one of which need to be kept to 
conform with the standard electroweak theory which has only 1 scalar 
Higgs field.  This is implemented in the FSM by imposing a condition:
\begin{equation}
\phi^{\tilde{2}}_r = - \epsilon_{rs} (\phi^{\tilde{1}}_s)^*,
\label{minframe}
\end{equation}
(where $\epsilon_{rs}$ is the 2-dimensional totally skew symbol)
which the frame vectors originally satisfy, and allows one of the 
framon fields, say $\bphi^{\tilde{2}}$, to be eliminated in terms of 
the other, say $\bphi^{\tilde{1}}$, thus minimizing again the number 
of new dynamical variables to be introduced by framing.  So, taken 
together with the observation in the preceding paragraph, the FSM 
can be said in this sense to be a minimally framed version of the 
SM.  (Notice that the parallel condition to (\ref{minframe}) in 
the colour framon cannot be similarly imposed because it would give 
to different components different physical dimensions \cite{dfsm}.)

With this last point clarified, one sees that the flavour framon in 
{\bf [FF]} above, with only one $\tilde{r}$ component kept as dynamical 
variable, say $\phi_r = \phi_r^{\tilde{1}}$, is just the standard 
Higgs field multiplied by a global factor $\balpha$, a triplet in 
$\widetilde{su}(3)$, which can be taken real in present applications, 
that is, a unit 3 vector in what will turn out for quarks and leptons 
to be 3-D generation space.  The quark and lepton mass matrices in 
FSM at tree level take on the common simple form:
\begin{equation}
m = m_T \balpha \balpha^\dagger     ,
\label{mfact}
\end{equation}
where $\balpha$, being a factor carried by the framon field, is the 
same for all quarks and leptons and only $m_T$, a positive real 
number, depends on the quark or lepton type, that is, whether up or 
down type quarks, or charged leptons or neutrinos.  Otherwise, the 
structure of the flavour sector is not much modified.

In contrast, the colour framon {\bf [CF]} represents new degrees of 
freedom with no analogue in the SM, and will lead to more structural 
changes.  First, it will change the structure of the vacuum.  We 
recall that the action of the framed theory, in particular the framon 
self-interaction potential $V$, has to be invariant under the double 
symmetry $G \times \tilde{G}$, which restricts $V$ to be of the form 
(with $\bphi = \bphi^{\tilde{1}}$, 
that is, the first column in the matrix $\Phi$, and $\bphi^{\tilde{a}} = (\phi^{\tilde{a}}_1, 
\phi^{\tilde{a}}_2, \phi^{\tilde{a}}_3)$ as the $\tilde{a}$ column of the matrix $\bPhi$):

\begin{eqnarray}
V & = & - \mu_W |\bphi|^2 + \lambda_W (|\bphi|^2)^2 \nonumber \\
  &   & - \mu_S \sum_{\tilde{a}} |\bphi^{\tilde{a}}|^2
        + \lambda_S \left( \sum_{\tilde{a}} |\bphi^{\tilde{a}}|^2 \right)^2
        + \kappa_S \sum_{\tilde{a} \tilde{b}} |\bphi^{\tilde{a}*}.\bphi^{\tilde{b}}|^2 
          \nonumber \\
  &   & + \nu_1 |\bphi|^2 \sum_{\tilde{a}}|\bphi^{\tilde{a}}|^2  
        - \nu_2 |\bphi|^2 \left( \sum_{\tilde{a}} \alpha^{\tilde{a}} 
          \bphi^{\tilde{a}}\right)^\dagger 
          \cdot \left( \sum_{\tilde{a}} \alpha^{\tilde{a}} \bphi^{\tilde{a}} \right),
\label{V}
\end{eqnarray}
up to quartic terms (for renormalizability), depending on 7 real coupling 
parameters, where one column of $\Phi$, say $\phi_r^{\tilde{2}}$, has 
already been eliminated leaving only $\phi_r = \phi_r^{\tilde{1}}$ as 
dynamical variable, as was explained in (\ref{minframe}) above.  Note in 
particular the $\nu_{1,2}$ terms in $V$ which link the flavour to the 
colour framon.  By choice, the 7 parameters as shown are all taken to be 
positive in the applications of the FSM so far implemented.  Minimizing
$V$, the vacuum is obtained, which gives a vacuum expectation value of 
the flavour $\Phi$ very similar to that of the Mexican hat potential of 
the standard electroweak theory, but a vacuum expectation value for the 
colour $\bPhi$ of the following form, in the local colour gauge where 
$\bPhi$ is hermitian and the global colour gauge where $\balpha =
(0, 0, 1)$:
\begin{equation}
\phivac \longrightarrow \zeta_S V_0 
   = \zeta_S \left( \begin{array}{ccc} Q & 0 & 0 \\
                                       0 & Q & 0 \\
                                       0 & 0 & P 
                            \end{array} \right),
\label{bPhivac}
\end{equation}
with:
\begin{eqnarray}
P & = & \sqrt{(1 + 2R)/3}, \\ 
Q & = &\sqrt{(1 - R)/3},
\label{PQdef}
\end{eqnarray}
where $\zeta_S$ is a quantity which describes the strength of the 
vacuum expectation value (vev) of $\bPhi$, and
\begin{equation}
R = \frac{\nu_2 \zeta_W^2}{2 \kappa_S \zeta_S^2}
\label{Rdef}
\end{equation}
can be thought of as the relative strength of the symmetry-breaking 
versus the symmetry-restoring term in dual colour $\widetilde{su}(3)$,
a ratio that will play an important role.

That the colour framon should acquire a non-zero vacuum expectation
value may seem at first 
sight disturbing, since the latter is often associated in our minds 
with the breaking of a local symmetry, while the colour symmetry here 
is supposed to be confining and to remain exact.  But this will not be 
so when one recalls 't~Hooft's alternative,
but mathematically equivalent, interpretation \cite{tHooft1,tHooft2} 
of the standard electroweak theory as a confining theory where local 
flavour remains exact, and only a global $su(2)$ symmetry associated 
with it (or dual to it) is broken.  So here, in analogy, colour can 
remain confining and unbroken, and that $\bPhi$ having a non-zero 
vacuum expectation value means 
only that the global dual colour symmetry $\widetilde{su}(3)$ is broken,
and since this last is supposed to play the role of generations for 
quarks and leptons, this scenario is exactly what one needs.

Indeed, when pushed further, this scenario, which we call the confining 
picture of 
't~Hooft, will lead to the result {\bf [HS]} noted above.  We recall that 
in the confining picture of the electroweak theory, the Higgs boson $h_W$ 
and the vector bosons $W^{\pm}, \gamma-Z^0$ appear respectively as the 
$s$- and $p$-wave bound states via flavour confinement of the Higgs 
scalar (here the flavour framon) with its conjugate, while quarks and 
leptons appear as bound states of the Higgs scalar (framon) with some 
fundamental fermion fields.  So now in the colour sector in FSM, there 
can be bound states via colour confinement of colour framon-antiframon 
pairs, in $s$-wave generically labelled $H$ and in $p$-wave generically 
labelled $G$, and of colour framons with fundamental fermion fields 
generically labelled $F$, in complete parallel respectively with $h_W$, 
$W^{\pm}, \gamma-Z^0$ and quarks and leptons of the electroweak theory, 
except with flavour and colour interchanged.  Further, these $H, G, F$ 
are predicted \cite{cfsm} to have little communication with the particles 
in the other set familar to us, hence the conclusion {\bf [HS]} above.
This also will be of relevance later. 
 
Further, whether in the symmetry-breaking or the confinement picture, 
by expanding framons about their vacuum expectation values, framon 
couplings can be worked out and renormalization effects by framon 
loops can next be studied.  Now, as already noted, framons differ from 
gauge bosons in that they carry two sets of indices, namely for the colour 
sector both in local colour $su(3)$ and in dual colour $\widetilde{su}(3)$.
This means that renormalization with framon loops will change not just the
normalization but also the orientation of the vacuum in dual colour space.  
Now of particular interest to the generation problem of quarks and leptons 
is the change in orientation (rotation) with respect to the scale $\mu$
of the vector $\balpha$ occurring 
in the mass matrix (\ref{mfact}).  Although $\balpha$, being a global 
quantity, is not renormalized by framon loops, it is coupled to the vacuum 
which is,
so that when the vacuum moves, it gets dragged along with it.  It is then 
of interest to find out how $\balpha$ rotates.

This question was studied in \cite{tfsm} to 1-framon-loop level, and from 
the renormalization group equation (RGE) 
derived, the following equations governing
the rotation of $\balpha$ result (where a dot denotes differentiation with 
respect to $t = 2 \ln \left( \frac{\mu}{1 \mathrm{Gev}}\right)$ for scale $\mu$):
\begin{equation}
\dot{R} = - \frac{3 \rho_S^2}{16 \pi^2} \frac{R(1-R)(1+2R)}{D}
          \left(4 + \frac{R}{2+R} - \frac{3 R \cos^2 \theta}{2+R}
          \right),
\label{Rdot}
\end{equation}
\begin{equation}
\dot{\theta} = - \frac{3 \rho_S^2}{32 \pi^2} \frac{R \cos \theta 
          \sin \theta}{D} \left(12 - \frac{6 R^2}{2+R} 
          - \frac{3k(1-R)(1+2R)}{2+R} \right),
\label{thetadot}
\end{equation}
and
\begin{equation}
\cos \theta \tan \phi = a,
\label{phi}
\end{equation}
with $a$ constant and 
\begin{equation}
D = R(1+2R) - 3R \cos^2 \theta + k(1-R)(1+2R),
\label{D}
\end{equation}
where $\theta$ and $\phi$ are polar co-ordinates of $\balpha$.  It is seen 
that the rotation of $\balpha$ is intimately coupled to the evolution with 
respect to the scale $\mu$ of the parameter $R$ noted in (\ref{Rdef}) before.  

These equations depend on 2 parameters, $\rho_S$ the coupling strength and 
$a$, an integration constant for an equation for $\phi$ already integrated
(\ref{phi}).
In solving the 2 coupled equations for 
$R$ in Eq. (\ref{Rdot}) and $\theta$ in Eq. (\ref{thetadot}), 2 more integration
constants are introduced, say $R_I$ and $\theta_I$.  Once knowing then the 
trajectory for $\balpha$, that is, not only the curve traced out by $\balpha$ 
on the unit sphere but also the speed with respect to $\mu$ at which it 
traces out this curve, the mass and mixing parameters of all quarks and 
leptons can be calculated using the general methodology developed earlier 
for phenomenological models with a rotating rank-one mass matrix (R2M2)
\cite{r2m2,bj} of form (\ref{mfact}).  To do so, one will need to feed in 
the values of the heaviest quark or lepton in each species, namely $m_t, 
m_b, m_\tau, m_{\nu_3}$, where the first 3 can be taken from experiment 
but the last, the Dirac mass of the heaviest neutrino, is unknown and has 
to be taken as another parameter.  In addition, one has to feed in the 
value of the strong-CP angle $\theta_{CP}$ and a value of a fudge factor 
$k$ to represent some secondary effects not yet calculable or calculated.
Hence, by adjusting the values of these 7 (real) parameter $\rho_S, a, 
\theta_I, R_I, m_{\nu_3}, \theta_{CP}, k$, one can fit the output values 
of the mass and mixing parameters of quarks and leptons to experiment.
This is done in \cite{tfsm} and Table \ref{tfsmfit} results.  This table 
includes all mass and mixing parameters of quarks and leptons and most 
are fitted to within the then (June 2014) experimental errors.  In effect, the FSM 
seems to have replaced some 17 of the SM's empirical parameters by only 
7.\footnote{Note that when this fit \cite{tfsm} was done, the situation with 
the CP-violating phase in the PMNS matrix for leptons was still unclear in 
experiment, nor had we understood it theoretically in the FSM, and the fit 
was done under the assumption that this phase was zero.  The effect of
this omission has 
since been investigated \cite{cpslept,cpslash}, yielding even a prediction for 
$\delta_{CP}$ for leptons which is within the present still rather loose 
experimental bounds.}

\begin{table}[h]
\begin{tabular}{|l|l|l|l|l|}
\hline
& Expt (June 2014) & FSM Calc & Agree to & Control Calc\\
\hline
&&&& \\
{\sl INPUT} &&&&\\
$m_c$ & $1.275 \pm 0.025$ GeV & $1.275$ GeV & $< 1 \sigma$&$1.2755$ GeV\\
$m_\mu$ & $0.10566$ GeV & $0.1054$ GeV & $0.2 \%$ & $0.1056$ GeV\\
$m_e$ & $0.511$ MeV &$0.513$ MeV & $0.4 \%$ &$0.518$ MeV\\
$|V_{us}|$ & $0.22534 \pm  0.00065$ & $0.22493$ & $< 1 \sigma$ &$0.22468$\\
$|V_{ub}|$ & $0.00351^{+0.00015}_{-0.00014}$& $0.00346$ & $< 1 \sigma$&$0.00346$ \\
$\sin^2 2\theta_{13}$ & $0.095 \pm 0.010$ & $0.101$ &$< 1 \sigma$ &$0.102$\\
\hline
&&&& \\
{\sl OUTPUT} &&&&\\
$m_s$ & $0.095 \pm 0.005$ GeV & $0.169$ GeV & QCD &$0.170$ GeV \\
& (at 2 GeV) &(at $m_s$) &running& \\
$m_u/m_d$ & $0.38$---$0.58$ & $0.56$ &  $< 1 \sigma$&$0.56$ \\
$|V_{ud}|$ &$0.97427 \pm 0.00015$ & $0.97437$ & $< 1 \sigma$&$0.97443$ \\
$|V_{cs}|$ &$0.97344\pm0.00016$ & $0.97350$ & $< 1 \sigma$&$0.97356$ \\
$|V_{tb}|$ &$0.999146^{+0.000021}_{-0.000046}$ & $0.99907$ &$1.65
\sigma$&$0.999075$ \\
$|V_{cd}|$ &$0.22520 \pm 0.00065$ & $0.22462$ & $< 1 \sigma$ &$0.22437$\\
$|V_{cb}|$ & $0.0412^{+0.0011}_{-0.0005}$ & $0.0429$ & $1.55 \sigma$&
$0.0429$ \\
$|V_{ts}|$ & $0.0404^{+0.0011}_{-0.0004}$ & $0.0413$ &$< 1 \sigma$& 
$0.0412$\\  
$|V_{td}|$ & $0.00867^{+0.00029}_{-0.00031}$ & $0.01223$ & 41 \% & $0.01221$\\
$|J|$ & $\left(2.96^{+0,20}_{-0.16} \right) \times 10^{-5}$ & $2.35
\times 10^{-5}$ & 20 \% &$2.34\times 10^{-5}$ \\
$\sin^2 2\theta_{12}$ & $0.857 \pm 0.024$ & $0.841$ &  $< 1 \sigma$& $0.840$\\ 
$\sin^2 2\theta_{23}$ & $>0.95$ & $0.89$ & $> 6 \%$ &$0.89$\\
\hline 
\end{tabular}
\caption{Calculated mass and mixing parameters of quarks and
leptons compared with experiment, quoted from \cite{tfsm}}
\label{tfsmfit}
\end{table}

\begin{figure}[h]
\centering
\includegraphics[scale=0.44]{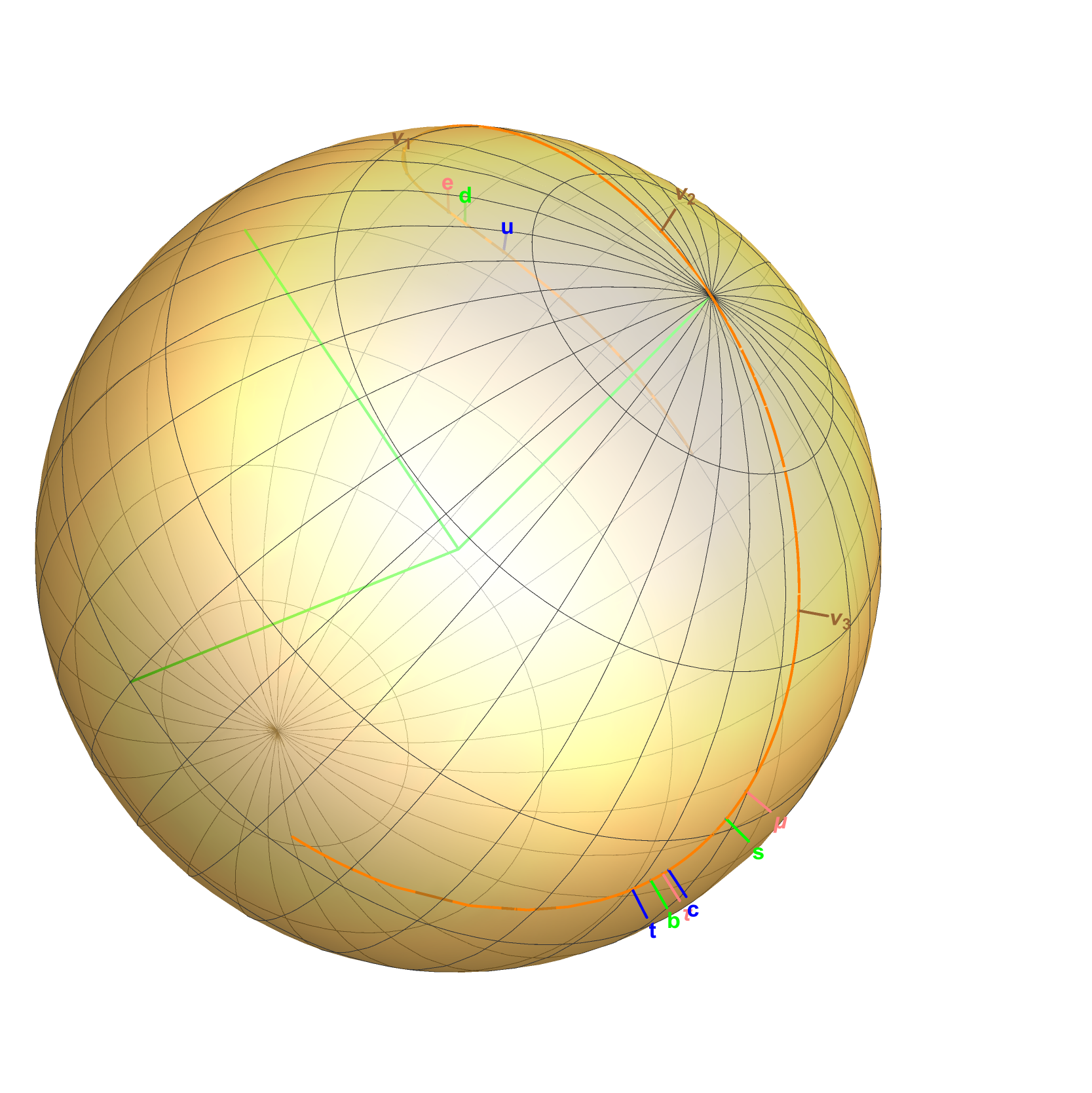}
\caption{The fitted trajectory for $\balpha$ on the unit sphere}
\label{Florosphere}
\end{figure}

\begin{figure}[h]
\centering
\includegraphics[scale=0.4]{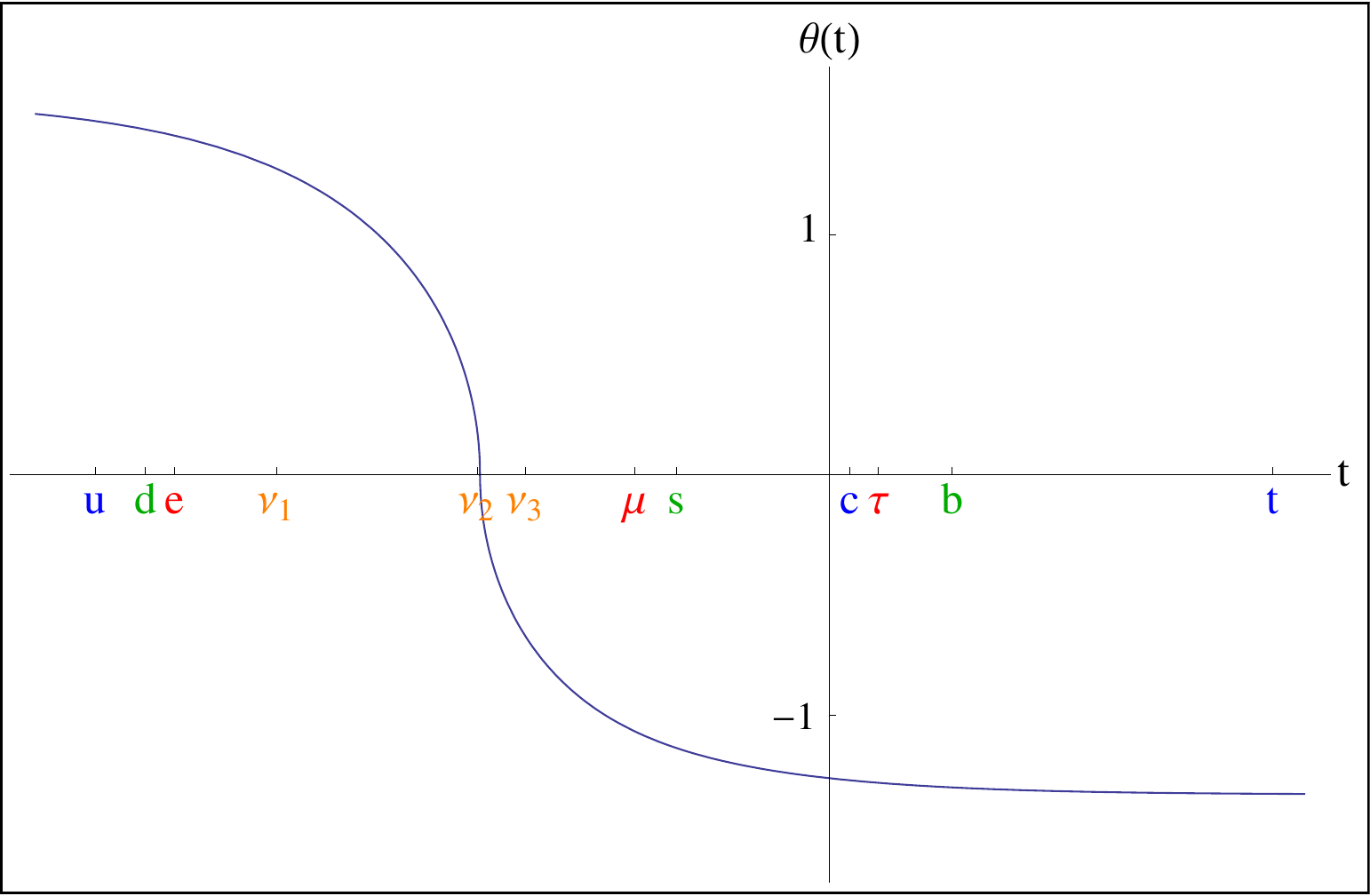}
\caption{Fitted $\theta$ as a function of $t = 2 \ln \left( \frac{\mu}{1 \mathrm{Gev}}\right)$}
\label{thetafit}
\end{figure}

\begin{figure}[h]
\centering
\includegraphics[scale=0.4]{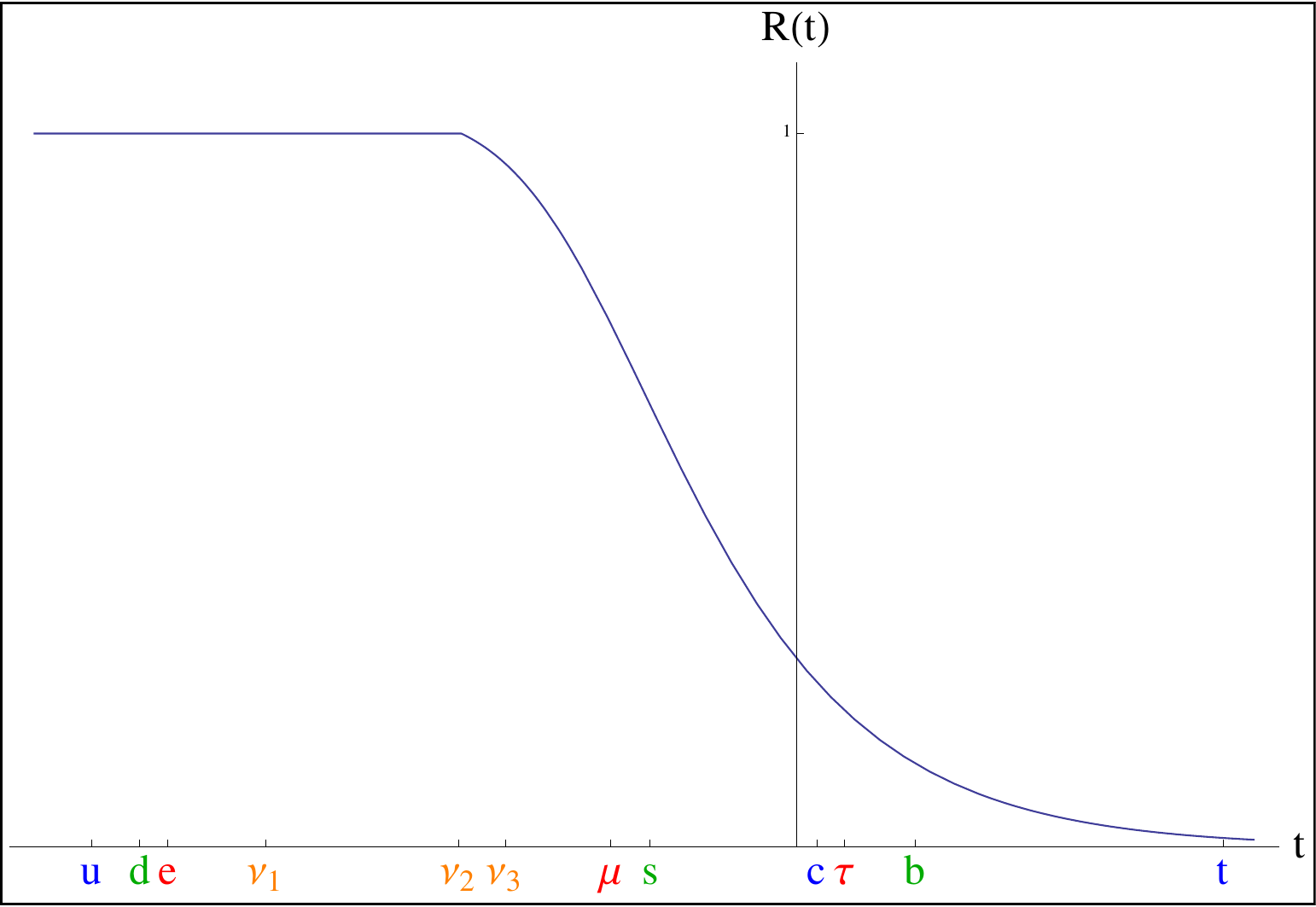}
\caption{Fitted $R$ as a function of $t = 2 \ln \left( \frac{\mu}{1 \mathrm{Gev}}\right)$}
\label{Rfit}
\end{figure}

The fitted trajectory of $\balpha$ on the unit sphere is shown in Figure
\ref{Florosphere} and the dependence of $\theta$ and $R$ on $t = 2 \ln \left( \frac{\mu}{1 \mathrm{Gev}}\right)$
are given in Figure \ref{thetafit} and Figure \ref{Rfit}.


\section{The vacuum transition at around 17 MeV}

For this paper, what interests us in the above fit is the surprising fact
that the FSM seems to have undergone a vacuum transition at a scale $\mu$ 
of around 17 MeV.  One sees in Figures \ref{thetafit} and \ref{Rfit} that 
at some point, say $\mu_T \sim 17\ {\rm MeV}$, the derivative of $\theta$ 
with respect to $\mu$ becomes infinite while $R=1$ and remains at 
$1$ for $\mu < \mu_T$.  

Now, the framon potential $V$ constructed, we recall, to be invariant under 
both the local symmetry $u(1) \times su(2) \times su(3)$ and the global 
symmetry $ \tilde{u}(1) \times \widetilde{su}(2) \times \widetilde{su}(3)$ 
as the FSM requires, allows both the following 2 classes of vacua as minima 
of $V$, one class with $R$ varying with changing $\balpha$, and another 
class with $R = 1$ and only $\balpha$ varying \cite{dfsm}.  The above then 
seems to say that as the scale lowers from high values where the vacuum is 
originally of the first class, there will come a point, say $\mu = \mu_T 
\sim 17\ {\rm MeV}$, where a transition will occur when the vacuum moves 
into the other class where $R$ takes the fixed value $1$. 

This transition, say VTR1, was first found in numerical solutions of the 
2 coupled equations (\ref{Rdot}) and (\ref{thetadot}) while effecting the 
fit, but can be seen also analytically from the 2 equations, as will be 
detailed in the Appendix.  It is also shown in the Appendix that the 
occurence of such a transition may not be restricted to the one-loop level 
at which the analysis has so far been carried out in detail, but seems to 
be kinematic in origin and may extend to higher loops.  The value $\mu_T$ 
at which the transition occurs will change with more loops, as will indeed 
the whole fit performed in \cite{tfsm}, but the occurence of such a 
transition seems likely to remain.  It has to be said though that neither 
the mathematical structure nor the physical significance of VTR1 has yet 
been fully understood.

Notice that, as is shown in the Appendix, this vacuum transition need 
not always appear in the form selected in \cite{tfsm}, but depends on 
the direction on the $\theta-R$ plane in which the transition point is 
approached.  However, that the fit to data in \cite{tfsm} should choose 
to approach VTR1 in the direction indicated in the analysis above is no 
mystery.  It was known already in R2M2 days \cite{r2m2} that in order 
to fit data the trajectory for $\balpha$ has to rotate at progressively 
greater speed as the scale $\mu$ decreases.  In this type of models, as 
now in FSM, the lower generations acquire masses and the mixing matrices 
acquire nonzero mixing only by virtue of the rotation of $\balpha$, and
the larger the values for the mass ratios, say $m_c/m_t, m_\mu/m_\tau$, 
or the mixing matrix elements, say $V_{cb}, U_{\mu 3}$, the faster the 
rotation will have to be.  Given then the empirical values:
\begin{equation}
m_c/m_t \sim 0.007, \ \ m_\mu/m_\tau \sim 0.06; \ \ |V_{cb}| \sim 0.04,
  \ \ |U_{\mu 3}| \sim 0.73    ,
\label{mratios}
\end{equation}
it follows that the curve of $\theta$ against $\mu$ in Figure \ref{thetafit}
has to grow progressively steeper as $\mu$ decreases, as is seen to be 
indeed the case in the fit.

That the curve has to grow progressively steeper as the scale lowers does
favour its approach to VTR1 in the direction it did, but cannot be taken 
as evidence that it should eventually acquire an infinite slope, namely 
that the transition VTR1 does indeed exist.  Can one find other empirical 
facts to support this latter claim?  The following two points are worth
mentioning.
\begin{itemize}
\item {\bf [E1]} 
First, empirically, from analyses in the SM, one wants $m_u < m_d$.  This is a bit of a puzzle because $m_t > m_b$, $m_c > m_s$, namely the U-type is heavier than the D-type in the 2 higher (that is, heavier) generations, but only $m_u < m_d$ in the lowest (that is, the lightest) generation.  Why?  The R2M2 type of models for quark masses did not solve the puzzle for, as explained above, the masses of the lower generations come, according to these models, from ``leakage'' from the highest generation via the rotation of $\balpha$.  Hence, the heavier the highest generation, the bigger will be the ``leak'' to the generations below, as confirmed by $m_t > m_b$, $m_c > m_s$.  But why $m_u < m_d$? It was only in \cite{tfsm} that we have for the first time chanced upon an answer for this puzzle in the FSM framework, and it comes about in the following fashion.  It was noted already some years ago \cite{features} that for the R2M2 type of models it is the geodesic curvature of the rotation trajectory of $\balpha$ that governs the lowest generation mass. Now in the trajectory given by the fit of \cite{tfsm}, as shown in Figure \ref{Florosphere}, the geodesic curvature changes sign when $\balpha$ passes through the transition point VTR1.  And this, as explained in detail in \cite{tfsm}, was what gave in that fit the very gratifying result $m_u < m_d$.  And this can be claimed, at least within the FSM framework, to be a consequence of the VTR1 transition.  We do not know, of course, whether $m_u < m_d$ has anything to do with $m_p < m_n$, the problem being nonperturbative is yet unsolved.  But if it does, then we might have here even an answer to an ancient question so crucial to our own existence.  

\item {\bf [E2]} The mass spectrum of the $H$ and $G$ states in the 
hidden sector {\bf [HS]} mentioned above can and has been calculated in 
\cite{cfsm} where it is shown that there is a bunch of states with mass 
eigenvalues $\zeta_S \sqrt{\third(1 - R)}$, where $\zeta_S$, we recall, is a 
measure of the vacuum expectation value
of the colour framon $\bPhi$ with an estimated value 
of about 2 TeV.  The physical masses of such states are to be measured 
by convention at the scale equal to their mass values, that is, solutions 
to the equation:
\begin{equation}
\mu = \zeta_S \sqrt{\third (1 - R(\mu))}.
\label{physmass}
\end{equation}
Hence, given that $(1 - R)$ is supposed to vanish at VTR1 at around 
17 MeV, these states must have a solution to (\ref{physmass}) for their 
physical mass just above but very close to 17 MeV.  Two of these states,
one scalar called $H_+$ and one vector called $G_3$, are known to have 
small couplings via mixing to our standard sector.  The first can be 
made to accommodate the $g - 2$ anomaly detected first some years ago 
at Brookhaven \cite{gminus2a} and confirmed, up to a point, two years ago 
at Fermilab \cite{gminus2b}, while the second can be made to accommodate 
the so-called Atomki anomaly detected in excited ${}^8Be$ decay by \cite{Atomki},
as was shown in \cite{fsmanom}.  
\end{itemize}
The first piece of evidence {\bf [E1]} is, of course, empirically solid 
but its relationship to VTR1 is not direct, while the relationship of 
the second piece {\bf [E2]} to VTR1 is direct, but the empirical facts 
as well as their FSM interpretations have yet to be further justified.
 
In brief, although in the FSM framework the transition VTR1 is strongly 
indicated by the mass and mixing patterns of quarks and leptons, one has 
as yet understood in depth and detail neither its mathematical structure 
nor its physical significance.  Nor has one gathered phenomenologically 
sufficient solid evidences as yet for its support.   However, this vacuum 
transition, if it indeed exists, would have potentially, we believe, such 
a significance with such far-reaching physical consequences that it is
worth being brought to the notice of the community even at this early 
stage of our own understanding of it.

In view of the importance of VTR1 to what follows, it would be  useful to have an estimate of the error on the transition scale 
given above as 17 MeV.  This, unfortunately, we are unable to  provide, because the fit given in \cite{tfsm} was only the best 
we could find by trial and error.  That we had chosen to proceed in such an ad hoc manner was not just negligence, there being a difficulty that we did not know how to overcome.  What we wanted was an overall description of the whole mass and mixing pattern of quarks of leptons, using all the data then available, but the different bits of data came with widely varying accuracy, with,for example, the electron mass known to $10^{-8}$, while for the lepton mixing angle $\sin \theta_{23}$ only a bound was known, so that a $\chi^2$ fit would make little sense without weighting and any weighting made would be arbitrary.  But if we allow ourselves to venture a guess, then from past experience and later applications 
\cite{KBslept}, we would suggest as an estimate of the error on the transition scale of about 1---2 MeV.  The reason for such a 
small-seeming error is the steepness of the curve for $\theta$ in Figure \ref{thetafit} near VTR1, so that a small change in 
$\mu$ would already make a big difference.  That is, of course, if we keep narrowly within the present FSM framework.    

We note in passing that although the exsistence of a vacuum transition at
17 MeV appears here as specific to the FSM via fitting to mass and mixing 
data, that there should be a new physical scale of that order was deduced 
by Bjorken \cite{bj} (7 MeV for him instead of 17 MeV here) via a line of 
dimensional arguments suggested by Zeldovich, and ultimately by Dirac, on 
much more general grounds.

\section{Implied metric in the hidden sector}

Our appreciation of the significance of the VTR1 transition is enhanced 
when considered in conjunction with the following facts.

The vierbeins in gravity are associated with a metric.  The framons 
in FSM, or at least their vev's, are supposed to be frame vectors 
too, like the vierbeins, and so should be associated with a metric 
as well.  In that case, at VTR1 when $Q$ vanishes, that metric may 
develop singularities and it would be interesting to find out what 
would ensue.  But first, of course, we have to clarify what is the 
metric implied by the framon vev's and what it measures.

Let us turn back and recall where this $\phivac$ actually turned 
up in FSM calculations.  It appeared in, for example, the Yukawa 
couplings, say (69) of \cite{cfsm}, in the combination:
\begin{equation}
\phivac^\dagger \psi_L(\sixth,2,3) = \chi_L(\half,2,1),
\label{chiL}
\end{equation}
where the numbers in brackets denote, respectively, the $U(1)$ charge, 
the dimension of the flavour representation, and the dimension of the 
colour representation.  This is one of the $F$'s in the hidden sector
{\bf [HS]}, being a colour-neutral bound state via colour confinement 
of the colour framon $\bPhi$, a colour triplet, with the fundamental 
fermion $\psi_L(\sixth,2,3)$, with the result still carrying the local 
flavour of the latter.  It was called a co-quark, being the parallel 
of a quark in the standard sector, only with the roles of flavour and 
colour interchanged.  

Now this $\chi_L$ is a vector in $\widetilde{su}(3)$ labelled by $\tilde{a}$ 
indices.  Suppose we ask how this vector is to be normalized.  To answer 
this, we first recall that $\psi_L$ in $su(3)$ colour is normalized thus:
\begin{equation}
\psi_L^\dagger \psi_L = 1.
\label{psinorm}
\end{equation} 
That being the case, then for the vector $\chi_L$ in dual colour 
$\widetilde{su}(3)$, just summing over the dual colour indices $\tilde{a}$ 
will not give us the norm, for according to (\ref{chiL}), this will give 
us:
\begin{equation}
\chi_L^\dagger \chi_L = \zeta_S^2 [Q^2 \psi_{L1}^* \psi_{L1} 
                                 + Q^2 \psi_{L2}^* \psi_{L2}
                                 + P^2 \psi_{L3}^* \psi_{L3}]
\label{chinormX}
\end{equation}
which is not unity.  However, if we introduce a metric, say,
\begin{equation}
{\bf G} = \zeta_S^{-2} \left( g^{\tilde{a} \tilde{b}} \right)
            = \zeta_S^{-2} {\rm diag}\,[Q^{-2}, Q^{-2}, P^{-2}],
\label{bG}
\end{equation}
then
\begin{equation}
|\chi_L|^2 = \sum_{\tilde{a} \tilde{b}} \chi_{L \tilde{a}}^\dagger 
                  g^{\tilde{a} \tilde{b}} \chi_{L \tilde{b}}
           = |\psi_L|^2 = 1.
\label{chinorm}
\end{equation}
It is here, then, that the metric associated with $\phivac$ comes in, 
namely to give the norms for the state vectors $\chi_L$ in the dual 
colour representation.

We notice that, contrary to the vierbeins in gravity where it is the 
global reference metric that is orthonormal (that is, Minkowskian) 
while the local metric is distorted, here it is the metric of the 
local (that is, dependent on space-time point $x$) symmetry which 
remains orthonormal while it is the global (independent of space-time 
point $x$) reference metric that is being distorted as per (\ref{bG}) 
above.  This is dictated by our requirement that local colour is 
confining so that its symmetry has to remain exact.

Although worked out specifically only for the co-quark (\ref{chiL}) 
above, the same analysis clearly applies to all the $F$s, namely all 
co-quarks and co-leptons, formed as bound states of the colour framon 
with fundamental fermion fields.  Their wave functions have all to be 
normalized with the metric (\ref{bG}). 

Indeed, the same assertion holds also for all the other particles in 
the hidden sector, namely vector and scalar states $G$ and $H$ formed 
by colour confinement from framon-antiframon pairs.  Let us work it 
out first for the $G$s which, in 't~Hooft's confinement picture, are 
$p$-wave bound states of framon-antiframon pairs:
\begin{equation}
G_\mu  =  \bPhi^\dagger D_\mu \bPhi.
\label{Gmu}
\end{equation}
To leading order:
\begin{eqnarray}
G_\mu & = & \phivac^\dagger \Omega^{-1} D_\mu \Omega 
            \phivac \\ \nonumber
      & = & \zeta_S \left( \begin{array}{ccc} Q & 0 & 0 \\
                                              0 & Q & 0 \\
                                              0 & 0 & P 
                    \end{array} \right)
            \tilde{C}_\mu 
            \zeta_S \left( \begin{array}{ccc} Q & 0 & 0 \\
                                              0 & Q & 0 \\
                                              0 & 0 & P 
                    \end{array} \right).
\label{Gmu1}
\end{eqnarray}
where:
\begin{equation}
\tilde{C}_\mu = \Omega^{-1} D_\mu \Omega 
\label{Cmutilde}
\end{equation}
is just a gauge transform of $C_\mu$ with $\Omega$ being the gauge 
rotation transforming from the standard gauge where the vacuum 
takes the form (\ref{bPhivac}) to a general local colour gauge. 

Now, for the colour gauge potential:
\begin{equation}
C_\mu = \sum_K C_\mu^K \lambda_K,
\label{Cexp}
\end{equation}
we can define the norm as (using the usual normalization for the lambda matrices):
\begin{equation}
\|C_\mu\|^2 = \half \Tr [C_\mu^\dagger C^\mu] 
           = \sum_K C_\mu^{K \dagger} C^{\mu K}.
\label{normC}
\end{equation}
 
Expanding similarly $\tilde{C}_\mu$ in (\ref{Gmu1}) above, one has:
\begin{equation}
G_\mu = \phivac \sum_K \tilde{C}_\mu^K \lambda_K \phivac.
\label{Gexp}
\end{equation}
One sees then that it will not do to take the norm of $G_\mu$ just 
as $\Tr [G_\mu^\dagger G^\mu]$.  One should instead insert the 
metric (\ref{bG}) when summing over tilde indices, thus:
\begin{eqnarray}
\|G_\mu\|^2 & = & \Tr [\bG G_\mu^\dagger \bG G^\mu] \\ \nonumber
           & = & \Tr [\bG \phivac \sum_K \tilde{C}_\mu^{K \dagger}
                      \lambda_K \phivac \bG \phivac
                      \sum_L \tilde{C}^{\mu L} \lambda_L \phivac]\\ \nonumber
            & = & \Tr [\phivac \bG \phivac 
                       \sum_{K,L} \tilde{C}_\mu^{K \dagger} \lambda_K
                       \phivac \bG \phivac
                       \tilde{C}^{\mu L} \lambda_L]\\ \nonumber
            & = & \sum_K \tilde{C}_\mu^{K \dagger} \tilde{C}^{\mu K},
\label{normG}
\end{eqnarray}
as claimed above, where in the last equality one has used the fact 
that:
\begin{equation}
\phivac \bG \phivac = 1. 
\end{equation}

The same is true also for the $H$ states which in 't~Hooft's language 
are $s$-wave bound states of the colour framon with its conjugate via 
colour confinement, thus $\bPhi^\dagger \bPhi$.  But, since $\bPhi$ 
has nonzero vev, it is the fluctuations about $\phivac$ that counts 
as the $H$s, thus $\phivac \delta \bPhi$.  Expanding then, one has:
\begin{equation}
\delta \bPhi = \left[ \sum_{L = 0,...,8} c_L \lambda_L \right] \phivac,
\label{Hexp}
\end{equation}
where $\lambda_0 = I/\sqrt{3}$.  Notice that since the $H$s are in 
$s$-wave, the expansion is in the hermitian basis $\lambda_L$, not 
the anti-hermitian basis $i \lambda_K$ for the $p$-wave states $G$.
Further, in our adopted notation, $\bPhi$, hence also $\delta \bPhi$
and $\phivac$, are matrices whose rows are labelled by colour 
and columns by dual colour, while $\lambda_L$ and $\lambda_K$ are
matrices whose rows and columns are both labelled by colour.  This 
gives the $H$ states.\footnote{Notice that these $H$ states obtained 
as framon-antiframon bound states as appropriate in the confinement 
picture here adopted differ in appearance from those exhibited in 
\cite{tfsm} which were worked out in the symmetry-breaking picture 
so as to conform with the earlier loop calculations that we had used.  
For this reason the norm also look different, but it can be and has 
been checked that the two representations are in fact equivalent.}
in parallel to (\ref{Gexp}) for $G$, as:
\begin{equation}
H = \sum_L c_L \phivac \lambda_L \phivac,
\label{Hexp1}
\end{equation}
where we recall from (\ref{bPhivac}) that $\phivac$ is a diagonal 
matrix with unequal diagonal elements.  Hence, as for $G$, the norm 
of $H$ is not given just by $\Tr [H^\dagger H]$ but by:
\begin{eqnarray}
\|H\|^2 & = & \Tr [\bG H^\dagger \bG H] \\ \nonumber
       & = & \sum_L |c_L|^2.
\label{normH}
\end{eqnarray} 
 
It appears then that the metric (\ref{bG}) applies to all particles 
$F, G, H$ comprising the hidden sector ${\bf [HS]}$ of the FSM when 
evaluating norms of their state vectors.\footnote{In an earlier 
paper \cite{dfsm}, when the existence of a hidden sector was not 
yet conceived, we had toyed with the idea that the metric (\ref{bG}) 
might apply to the standard sector.  We now think that that was 
wrong, and that the metric (\ref{bG}) should apply only to vectors 
in the hidden sector as argued above.}  Recall now that as the scale 
lowers towards VTR1, $R \rightarrow 1$, and 2 entries in (\ref{bG}) 
blow up, although since the corresponding components in the relevant 
state vectors vanish, there are no actual infinities in the norms.  
Nevertheless, this behaviour of the metric at VTR1 seems likely to 
have significant effects on the physics in the hidden sector where 
it operates.  Our knowledge and understanding of the hidden sector 
being so meagre at present, our imagination has so far failed us in 
finding detectable effects therein, but we have noted already two 
interesting reflections {\bf [E1], [E2]} they have on the standard 
sector.  In any case, since the dynamics of the hidden and standard 
sectors are intimately connected\footnote{Recall that even the rotation of 
$\balpha$ which is supposed to give rise to the hierarchical masses of 
quarks and leptons comes from renormalization in the hidden sector.}, 
the noted singularity at VTR1 in the metric (\ref{bG}) in the hidden 
sector does not concern just that sector alone but could have grave 
repercussions on the whole scheme and has thus to be understood.

\section{Kaluza-Klein embedding} 

Having failed so far in finding a verifiable effect of the transition 
VTR1 in the hidden sector itself, yet still wanting to find an example 
of such, we have then let our imagination loose, and it flew off at a 
tangent in the following direction.  In gravity the vierbeins lead not 
just to a metric measuring the norms of vectors but to an actual metric 
of space-time itself.  Attempting to mimic that, we try to imagine the 
whole FSM scheme being embedded, in Kaluza-Klein (KK) fashion, into a 
higher-dimensional space-time, a scheme we may call KFSM for short.  In 
that case, we imagine, the vacuum transition VTR1 might lead to unusual 
behaviour of the metric in space-time itself and give rise to observable 
cosmological effects.

Before pushing further, however, two remarks are here appropriate.  
First, in making the KK assumption suggested above, one has gone 
beyond the original remits of the FSM.  Hence, whatever new results 
one will obtain in this new venture, whether good or bad, should not 
affect the results obtained before with the FSM for particle physics 
to which it has so far been limited.  And, of course, nor can the FSM 
alone claim credit for whatever the new venture might hope to achieve.   

Secondly, we recognize, of course, that even the embedding of the SM 
into a KK theory is a sophisticated topic in itself already worked on 
by many eminent physicists without so far a definitive solution, and 
we have nothing to add to that.  The embedding of the FSM into KK is 
therefore even more of an open problem and may not even be possible.  
We shall show, however, that so long as the KK embedding somehow goes 
through, what concern us in what follows are just a few compactified 
dimensions picked out by framing in the FSM as requiring our special 
attention, and so may not depend on details of how the KK embedding 
is actually effected.

With these reservations understood, let us then embark on our new 
adventure into the unknown, and prepare ourselves for confrontation 
with some unexpected consequences.

In a Kaluza-Klein theory, the compactified dimensions representing 
the gauged internal symmetry are generally supposed to be small in 
size so as to have escaped present detection, but it is usually not 
specified what actual sizes they do have.  For example, some would 
suggest the Planck length.  However, when the gauge theory involved 
is framed, as what interests us here, two significant changes ensue:
\begin{itemize}
\item The framons, though frame vectors in internal symmetry space, 
are Lorentz scalars in ordinary 4-dimensional space-time and so, 
as such, may acquire non-zero vacuum expectation values depending on 
their self-interaction potential.
\item And, being frame vectors in internal space, those framons with 
non-zero vev's would imply non-zero components of the metric, each 
of a certain size prescribed by the self-interaction potential in 
some specified direction of the compactified space representing the 
internal symmetry.
\end{itemize}
In other words, when the gauge theory embedded in the Kaluza-Klein 
scenario is framed, the physical concept of length is introduced 
into the compactified dimensions which they did not seem to have 
possessed before.  One is then not free to assign arbitrary small 
sizes to all the compactified dimensions.  Some of them will have 
to acquire the nonzero physical sizes specified and assigned to them 
by the framon self-interacting potential.  We shall refer to these 
latter as the vev-enhanced dimensions, as opposed to those remaining 
dimensions corresponding to framon components with zero vev's with 
therefore presumably still their original minimal size.

Specialize now to the FSM where the SM gauge symmetry is $G = u(1) \times 
su(2) \times su(3)$ and the flavour and colour symmetries are framed.  
In the confinement picture 
of 't Hooft which the FSM has adopted and extended, both the global 
flavour and colour 
symmetries are broken and both the flavour and colour framons acquire 
non-zero vev's.  Let us first work it out for colour, which is of special 
interest to us here.  For this, as already shown above, the framon has 
the vev (\ref{bPhivac}), and this has led to the metric (\ref{bG}) for 
measuring lengths of vectors in the hidden sector.

Notice that $Q$ and $P$ are dimensionless but $\zeta_S^{-1}$ has the 
physical dimension $L$ of length.  Hence (\ref{bG}) means that when 
applied to vectors in the colour symmetry space, which has initially 
no physical dimension, they will now acquire a norm with the physical 
dimension of length.  Equivalently, which may be more convenient, one 
can shift the physical dimension to the vectors, saying that they 
are measured in units of $\zeta_S^{-1}$ with dimension $L$, and that 
their norms are to be defined with the dimensionless metric:
\begin{equation}
\hat{\bf g} =  (g^{\tilde{a} \tilde{b}})
        =  \left( \begin{array}{ccc} Q^{-2} & 0 & 0 \\
                                      0 & Q^{-2} & 0 \\
                                      0 & 0 & P^{-2} \end{array}
            \right).
\label{bgup}
\end{equation}

Imagine next that there is a field of such vectors over those 
compactified dimensions which correspond to colour in the 
Kaluza-Klein scenario, and that this metric measuring the norms of these 
vectors is then to be taken as the contravariant metric.  To this, 
the covariant metric measuring distance in the underlying 
Kaluza-Klein space would be the inverse:
\begin{equation}
\check{\bf g} =  (g_{\tilde{a} \tilde{b}}) 
         =  \left( \begin{array}{ccc} Q^2 & 0 & 0 \\
                                      0 & Q^2 & 0 \\
                                      0 & 0 & P^2 \end{array}
            \right),
\label{bgdown}
\end{equation}
giving in those dimensions of the KK space the elemental distance 
$d\bsigma$ as:
\begin{equation}
d\bsigma^2 = g_{\tilde{a} \tilde{b}} dx^{\tilde{a}} dx^{\tilde{b}}.
\label{dsigmasq}
\end{equation}
This metric (\ref{bgdown}) is also dimensionless, in keeping with 
the convention adopted above for (\ref{bgup}).  What is the physical 
dimension then of $d\bsigma$?

This distance $d\bsigma$ is measured in the manifold of compactified 
dimensions corresponding to colour in the Kaluza-Klein space while 
the vectors with norms of the dimension of length $L$ above live in 
the tangent space to this manifold, but locally, the tangent space 
and the manifold coincide.  It seems thus natural that the line 
element $d\bsigma$ should have the same physical dimension as that 
of the vectors above, namely $L$ also.  This means that, since we
have adopted the covention that both the metrics (\ref{bgup}) and 
(\ref{bgdown}) are dimensionless, it is the displacements in 
co-ordinates $dx^{\tilde{a}}$ which should carry this physical 
dimension $L$.  This is exactly what is needed if these are to be 
taken as compactified dimensions of space as KK suggest.

This last assertion is by itself quite intriguing.  In an ordinary 
KK theory when a gauge symmetry is embedded as extra dimensions in 
space-time, the symmetry space has in itself no physical dimension.  
But if it were to play the role of compactified dimensions in space 
as KK suggests, then it has to be given the physical dimension of 
$L^d$, where $d$ is the number of the extra dimensions introduced.  
Usually in the literature, as already noted above, this needed 
physical dimension is introduced via the Planck constant, tacitly 
thus assuming that the compactified dimensions are of the order of 
the Planck length in size.  Here, we seem to be saying that in FSM, 
because the gauge symmetry is framed and its global dual is broken, 
some dimensions in that symmetry space get given not only a size 
but a size in the appropriate dimension of length $L$, and this size 
is of order $\zeta_S^{-1}$, with $\zeta_S$ being the strength of the 
vacuum expectation value of the framon field, which is much larger 
than the Planck length.  That such a result should obtain 
in the FSM is traced to the confinement picture of 't~Hooft 
(extended to include colour) which 
gives the state vector of a co-quark as $\chi_L$ in (\ref{chiL}),
namely as a bound state of the fundamental fermion field with a 
framon, thereby making it carry the physical dimension of the latter.

Explicitly then, from (\ref{dsigmasq}) and (\ref{bgdown}), we have:
\begin{equation}
d\bsigma^2 = Q^2 (dx^{\tilde{1}})^2 + Q^2 (dx^{\tilde{2}})^2 
            + P^2 (dx^{\tilde{3}})^2,
\label{dbsigmasqexp}
\end{equation}
in the local $su(3)$ gauge where the framon $\bPhi$ is hermitian 
and the global $\widetilde{su}(3)$ gauge where $\balpha$ points in 
the $\tilde{3}$ direction.  We note also that $Q$ and $P$, though dependent 
on the renormalization scales $\mu$, do not depend on 
the co-ordinates $x^{\tilde{a}}$.  This metric then signifies a 
3-torus in space at every $\mu$.

This means then that of the compactified dimensions of the space 
representing the  colour symmetry in the KK theory, those 3 
denoted by the diagonal elements of $\check {\bf g}$ stand out as 
having acquired by symmetry-breaking a non-minimal size of order 
$\zeta_S^{-1}$, these forming together a 3-torus subspace of the 
colour symmetry space, while the other dimensions therein 
remain of the original minimal size.

A similar analyis can be carried out for the flavour symmetry
\cite{zmixed,cfsm} which gives the line element in that subspace as:
\begin{equation}
d\sigma^2 = \zeta_W^{-2} [\half (dx^{\tilde{1}})^2 
            + \half (dx^{\tilde{2}})^2],
\label{dsigmasqF}
\end{equation}
again signifying that of the several compactified dimensions of the 
space representing the flavour symmery in the KK theory, 
those 2 denoted by the diagonal elements of:
\begin{equation}
\check{\rm g} = \zeta_W^{-2} \,{\rm diag}\,[\half, \half],
\label{gdown}
\end{equation}
stand out as having acquired by symmetry-breaking a non-minimal 
size of order $\zeta_W^{-1}$ and form a 2-torus subspace of the 
flavour symmetry space, while the other dimensions remain of the 
original minimal size.

Altogether then, one seems to have acquired, by symmetry-breaking,
5 what we might call vev-enhanced dimensions of the 
internal symmetry space each with a size prescribed by the framon 
potential while the remaining dimensions of that space presumably
would retain their original minimal size.  And from the way this 
conclusion was worked out above, it seems that the world would 
look rather different to the standard or to the hidden sector 
particles.  

To the standard sector particles such as the quarks, the leptons, 
the vector bosons $W-Z$ and the Higgs boson $h_W$, space-time would 
consist of (i) the usual 4 dimensions plus (ii) a vev-enhanced 
2-torus of size of order $\zeta_W^{-1}$ as measured by the metric 
(\ref{gdown}), plus (iii) some non-vev-enhanced dimensions with 
still unspecified minimal size corresponding to the remaining 
components of the flavour framon $\Phi$ with zero vev's, while 
(iv) the colour dimensions and (v) the original $u(1)$ would appear 
in the usual Kaluza-Klein fashion as gauged symmetry spaces for 
colour and electromagnetism.  

To the hidden sector particles $F$, $G$, and $H$, on the other hand, 
space-time would consist of (i) the usual 4 dimensions plus (ii) a 
vev-enhanced 3-torus of size of order $\zeta_S^{-1}$ as measured by 
the metric (\ref{bgdown}), plus (iii) the non-vev-enhanced dimensions 
with still unspecified minimal size corresponding to the remaining 
components of the colour framon with zero vev's, while (iv) the 
flavour dimensions and (v) the original $u(1)$ would appear in usual 
Kaluza-Klein fashion as the gauged symmetry spaces for flavour and 
electromagnetism.  

To the photon, however, which is in the FSM a mixture of a $u(1)$ 
gauged particle with the standard sector vector boson $Z$ and with a
 hidden sector vector boson $G$ \cite{zmixed,cfsm}, the world will 
look again different in a way which will be of particular interest 
to us.  To the photon, space-time would consist of (i) the usual 4 
dimensions plus (ii) a 2-torus of size $\zeta_W^{-1}$ plus (iii) a 
3-torus of size $\zeta_S^{-1}$ plus (iv) other compactified dimensions 
of minimal size, while (v) the original $U(1)$ dimension would remain 
in the KK fashion as the gauged symmetry space of electromagnetism.

Our next question is how this structure will change with changing 
scale.  In particular, we are interested in how it would behave as 
$\mu$ approaches the VTR1 transition, say, from above.  Now, of the 
many compactified dimensions with minimal sizes we know nothing, 
and can only assume that they remain more or less oblivious to such
scale-changes.  Of the 5 vev-enhanced dimensions, the 2 from flavour 
have a size proportional to $\zeta_W^{-1}$, where $\zeta_W 
\sim 246$ Gev and is known to be only weakly dependent on scale.  
But the remaining 3 from colour have different sizes: $\zeta_S^{-1}$ 
times $Q$ or $P$, as given in ({\ref{PQdef}), where $R$ is known to 
be strongly scale-dependent, as seen in Figure \ref{Rfit}.  Indeed, as 
the scale approaches the vacuum transition VTR1 at around 17 MeV, we 
see that those 2 dimensions with size proportional to $Q$ would shrink
to zero.  This means that the volume of the whole 3-torus formed from 
the 3 vev-enhanced dimensions would collapse to zero altogether.

From the fit in \cite{tfsm} to the mass and mixing data of quarks and 
leptons, one has in FSM even a rough picture of how this collapse may 
behave as it nears the VTR1 transition point.  According to the metric 
(\ref{bgdown}), the 3-torus in the vev-enhanced dimensions has a volume 
proportional to:
\begin{equation}
\zeta_S^{-3} (1 - R) \sqrt{1 +2R}.
\label{shrink}
\end{equation}  
The behaviour of $R$ near VTR1 is given in Figure \ref{Rfit}, but there 
is as yet no direct information on $\zeta_S$ except for an estimated 
value of $\zeta_S \sim 2$ TeV at high scales (say around the top mass 
$m_t$) from \cite{zmixed}.  However, we recall from (\ref{Rdef}) that:
\begin{equation}
\zeta_S^{-1} = \sqrt{R \frac{2 \kappa}{\nu_2 \zeta_W^2}}
\label{zetaSinR}
\end{equation}  
where $\zeta_W$ is known to be about 246 GeV and weakly dependent on 
scale, while $\kappa$ and $\nu_2$ are dimensionless couplings in the 
framon potential (\ref{V}) which are naturally of order unity and 
hence also likely to be weakly scale-dependent.  Though crude, this 
last assertion is checked to be consistent with the above estimate 
of $\zeta_S \sim 2$ TeV at $\mu \sim m_t$ where $R \sim 0.01$.  If 
so, then we have roughly 
that $\zeta_S^{-1} \sim \sqrt{R}$ in 
scale-dependence giving instead of (\ref{shrink}):
\begin{equation}
R^{3/2} (1 - R) \sqrt{1 + 2R},
\label{shrink1}
\end{equation}
for the scale-dependence of the volume of the vev-enhanced 3-torus 
in colour near the VTR1 transition.  And, if we ignore the unkown 
scale-dependence of the non-vev-enhanced Planck sized dimensions, 
the same behaviour would apply to the whole compactified subspace 
corresponding to the colour symmetry in the KK scenario.

From Figure \ref{Rfit} then, one obtains via (\ref{shrink1})  
Figure \ref{howshrink}.  The volume of the subspace corresponding 
to colour is seen to start off at high $\mu$ ($\sim m_t$) with a 
small value, rising at first as $\mu$ lowers, in line with the 
usual 3 dimensions of space, but then as the scale nears the VTR1 
transition at $\mu_T \sim$ 17 MeV, it drops quite abruptly to 
zero.  

\begin{figure}[h]
\centering
\includegraphics[scale=0.4]{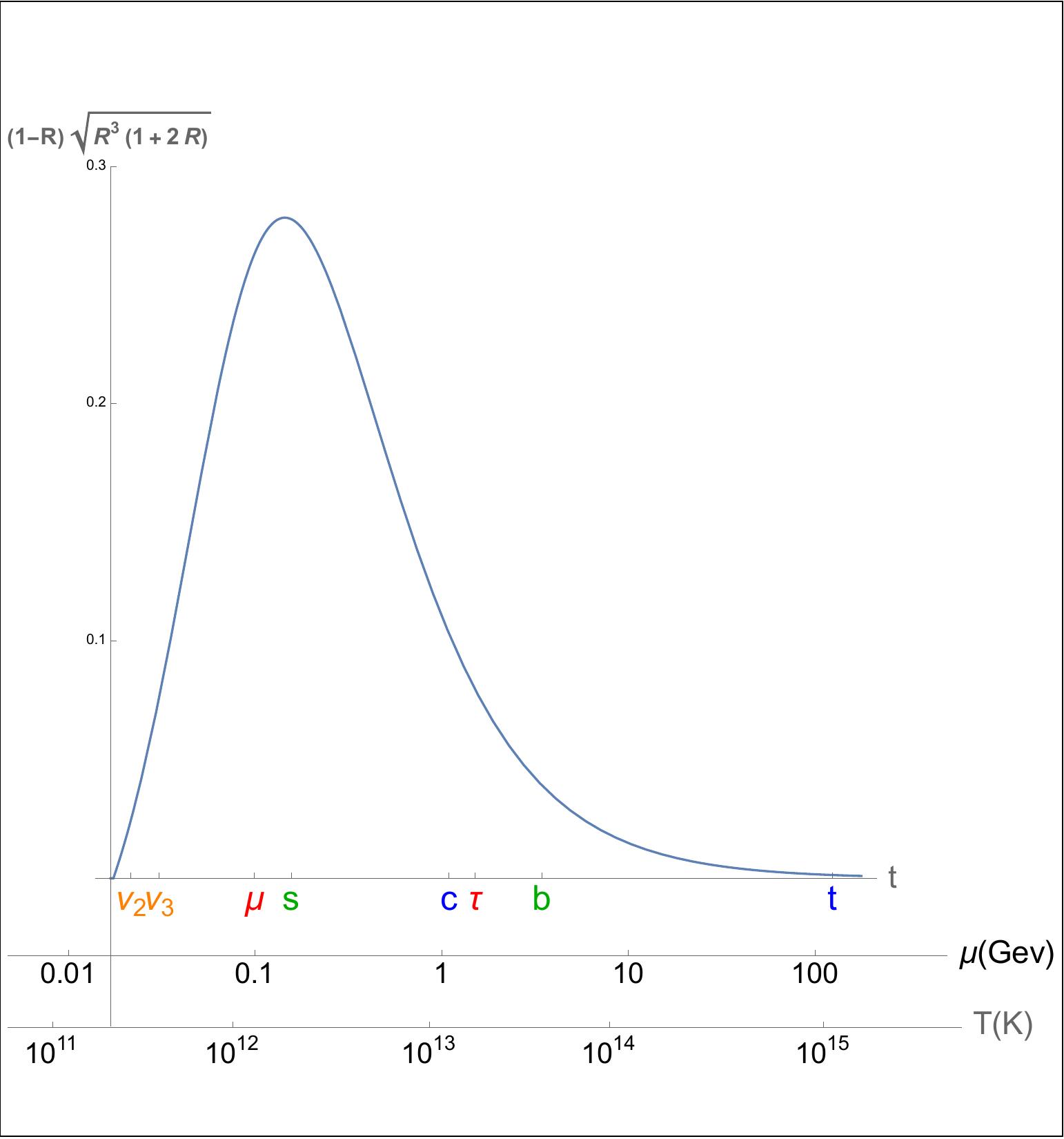}
\caption{A sketch of how, as surmised, the volume of the compactified 
colour dimensions would vary with scale and shrink eventually to zero 
as the scale lowers towards the VTR1 transition.}
\label{howshrink}
\end{figure} 

In other words, the KFSM scheme seems to say that, driven by some 
microscopic dynamics mostly in the hidden sector, space-time in the 
Kaluza-Klein scenario would lose at VTR1 quite abruptly 2 of its 
compactified spatial dimensions.  If this is true, dramatic effects,
it seems, can hardly be avoided.  In the next section, we aim to 
set our imagination loose and try to figure out how the evolution 
of the early universe might be affected---purely in a spirit of 
exploration on our part, however inadequately equipped we know we 
really are for the task.

\section{Possible effect on the horizon problem}    

Suppose we accept the above picture of the FSM embedded Kaluza-Klein 
fashion into space-time, namely 
that space-time is effecively 10-dimensional: 4 for the usual, plus 1 
compactified for the $u(1)$ gauge, plus 2 compactified vev-enhanced for 
flavour, plus further 3 compactified vev-enhanced for colour.  The 2 
vev-enhanced dimensions for flavour have a size of order $\zeta_W^{-1}$ 
where $\zeta_W$ is the vev of the standard Higgs field of order 246 GeV, 
while the 3 vev-enhanced dimensions for colour have sizes of order 
$\zeta_S^{-1}$ where $\zeta_S$ is the vev strength of the colour framon 
(a new ingredient of FSM with no analogue in the SM) thought to be of 
order TeV at high scales but probably quite strongly scale-dependent.  
There are other compactified dimensions with minimal sizes about which 
we know nothing and can and shall largely ignore.  Let us accept also, 
as put forward 
above, that because of the vacuum transition VTR1 found in 
FSM at a scale of around 17 MeV \cite{tfsm}, 2 of the 3 vev-enhanced 
dimensions for colour shrink to zero as the scale approaches VTR1 from 
above, as depicted in Figure \ref{howshrink}. 

Suppose further that as the universe cooled, it was its temperature at 
every stage which set the scale so that we can follow the development 
of VTR1 in Figure \ref{howshrink} just by converting the scale $\mu$ 
to the temperature $T$ as shown.  One sees then that as the universe 
cooled from some initial high temperature, the compactified dimensions 
for colour first expanded, in line with the usual 3 space dimensions, 
but then started to contract when the transition VTR1 at a scale of 
about 17 MeV was approached.  This corresponds to a temperature of 
about $2.0 \times 10^{11}$ K when the universe was around $2.5 \times 
10^{-3}$ s old. 

What would happen then at this transition when 2 of the compatified 
dimensions are supposed to shrink to zero?  This reminds one of a 
balloon being squeezed---if compressed in one direction, it would 
expand in the other directions, as long as the amount of air it contains is 
kept the same.  Would something similar occur to the early universe 
at VTR1? 

It would seem that it might.  The universe around the time of the VTR1 
was radiation-dominated, so that in terms of the balloon analogy that 
we cavalierly suggested above, the air inside the balloon would here 
be replaced by black body radiation.  This in a sense makes the problem 
easier, since the energy density per unit volume would, by Jean's Law,
be a function of just the temperature.  In Figure \ref{howshrink} we 
saw that VTR1 is supposed to occur over a relatively short range of 
temperature.  Suppose that as a first approximation we take the 
transition to be isothermal, then the energy density of the black 
body radiation contained would remain the same.  And if the number of 
photons also remains the same, the volume they occupy should remain 
the same before and after the transition.  Hence at VTR1, when 2 of 
the compactified dimensions shrink to zero, the other dimensions will 
have to expand to compensate, as envisaged in the balloon analogue.
 
In the present Kaluza-Klein scenario many dimensions are involved.  
Apart from the usual 4 dimensions of space-time, we have in addition 
the compactified dimensions corresponding to the embedded internal 
symmetry $G$.  Hence, when 2 of the latter dimensions corresponding 
to colour are squeezed by VTR1 as proposed, we have to ask which of
the remaining dimensions will expand to compensate.  We recall first 
from (\ref{bgdown}) that when $R \rightarrow 1$ as VTR1 is approached
and 2 diagonal entries in $\check{\bf g}$ shrink to zero, the third 
entry just goes gently to $1 \times \zeta_S^{-2}$ and cannot expand.  
Neither can the compactified dimensions corresponding to flavour, the 
size of which were seen to be governed by $\zeta_W$, known in the 
standard electroweak theory to be only weakly scale-dependent.  Of 
the compactified dimensions there remain then only the $u(1)$ gauge 
and the ones we labelled as non-vev-enhanced all of minimal Planck 
size which we have assumed, and shall continue to assume, to be 
oblivious to scale changes.  In that case, it would seem that at VTR1 
when the 2 collapsing dimensions shrink to zero, only the 3 standard 
spatial dimensons can freely expand to compensate for the shrinkage.

Supposing this is true, let us ask next by how much the 3 spatial 
dimensions might be expected to expand under VTR1.  The 2 collapsing 
dimensions are supposed to have started just before the VTR1 with a 
value of order $\zeta_S^{-1} \sim \zeta_W^{-1}$.  If they actually 
shrank to zero after VTR1, then the expansion of course would be 
infinite.  Let us however take zero practically to mean only the 
smallest length known, namely the Planck length:  
\begin{equation}
\ell_{Pl} = m_{Pl}^{-1} \sim 1.616 \times 10^{-35}\ {\rm m},
\label{Plancklength}
\end{equation}
then the expansion will be large, but finite.

Assuming next that the universe remained isotropic, for example with 
the Robertson-Walker metric \cite{RW1a,RW1b,RW1c,RW2}, and remaining still for the 
moment in the approximation when VTR1 is taken isothermal so that 
the volume remained the same before and after the transition, we will
then have:
\begin{equation}
{\cal R}_F^3 \ell_{Pl}^2 \sim {\cal R}_I^3 {\ell_I}^2,
\label{Exest1}
\end{equation}
where ${\cal R}_I$ is initial size of the universe before the VRT1
transition, ${\cal R}_F$ the expanded size after the transition, and 
$\ell_I$ is the initial size of the 2 compactified colour dimensions 
which were shrunk to the size of the Planck length by the transition.

Suppose we substitute then for $\ell_I$ the above order of magnitute 
estimate:
\begin{equation}
\ell_I \sim \zeta_S^{-1}(\mu \sim \mu_T) \sim \zeta_W^{-1} 
   \sim (246\ {\rm GeV})^{-1} \sim 8.0 \times 10^{-19}\ {\rm m}.
\label{ellI}
\end{equation}
We obtain the order of magnitude estimate for the expansion factor:
\begin{equation} 
{\cal E}_{\rm isotherm} = \frac{{\cal R}_F}{{\cal R}_I} 
   \Bigg\vert_{\rm isotherm}
\label{calE}
\end{equation}
as:
\begin{equation}
{\cal E}_{\rm isotherm} \sim \left( \frac{\ell_I}{\ell_{\rm Pl}} \right)^{2/3}
   = \left(\frac{8.0 \times 10^{-19}}{1.6 \times 10^{-35}} \right)^{2/3} 
   = 1.4 \times 10^{11}.  
\label{Exest2}
\end{equation}
That such an expansion would be enough to imply a universe smaller in size 
than the horizon before the transition can be seen as follows.

The portion of the universe that is observable to us, or in other 
words the horizon at present, is 
\begin{equation}
t_0 = 1.38 \times 10^{10}\ {\rm yr},
\label{t0}
\end{equation}
which in units of length is
\begin{equation}
{\cal R}_0 = {\cal H}_0 \sim 3 t_0 \sim 4.14 \times 10^{10} \ {\rm ly} 
           \sim 3.92 \times 10^{26}\ {\rm m}.  
\label{calR0}
\end{equation}
For general time $t$, however, we need to distinguish between the 
two concepts: ${\cal R}$ the size of the universe, and ${\cal H}$ 
the horizon, since the two evolve with time differently. 
 
Tracing the evolution backwards in the standard Big Bang model, one 
estimates that the part of the universe we now observe, or the 
universe for short, was, at its age of $2.5 \times 10^{-3}$ s just 
after the VTR1 transition, only about 
\begin{equation}
{\cal R}_{T+} \sim 3.7 \times 10^{15}\ {\rm m} 
\label{calRT+}
\end{equation}
in size.  
In arriving at this estimate, we 
have taken account of the knowledge \cite{weinbergcos,pdg} that the 
energy density of the universe was dominated by matter from now at 
an age of $1.38 \times 10^{10}$ yr back to about $4 \times 10^4$ yr, 
when its size depended on time roughly as $t^{2/3}$, but from 
then onwards back to its age just after VTR1, when it was radiation 
dominated, and its size depended on time as $t^{1/2}$.  

On the other hand, the horizon ${\cal H}$ depended on time roughly 
as $2 t$ when the universe was radiation dominated, so 
that at the time around VTR1, when its age was $\sim 2.5 \times 
10^{-3}\ {\rm s}$, the horizon was only of order:
\begin{equation}
{\cal H}_{\rm VT} \sim 1.5 \times 10^{6}\ {\rm m}.
\label{HT}
\end{equation}
In other words, the 
universe then of size (\ref{calRT+}), as estimated above, was made 
up of numerous causally disconnected bits, as it was already at the 
decoupling time when the CMB was formed, only exacerbated. 

However, given the size of the universe ${\cal R}_{T+}$ in (\ref{calRT+})
just after the transition and the estimate (\ref{Exest2}) for the 
expansion factor due to VTR1, one concludes that the universe would 
have expanded from a size ${\cal R}_{T-}$ just before the transition of 
only:
\begin{equation}
{\cal R}_{T-} \sim \frac{3.7 \times 10^{15}\ {\rm m}}{1.4 \times 10^{11}}
              \sim  2.7 \times 10^4\ {\rm m},
\label{calRT-}
\end{equation}
which would fit comfortably inside the horizon at that time, 
namely the ratio of size to horizon: 
\begin{equation}
\frac{{\cal R}}{{\cal H}}\Bigg\vert_{T = 17\ {\rm MeV}}
   \sim  0.018; \ \ \ {\rm isothermal\ approx.}
\label{Rratioiso}
\end{equation}

This estimate is inaccurate, however, having been derived assuming 
that the VTR1 transition is isothermal, which it clearly is not.  
As seen in Figure \ref{howshrink}, it seems to have occurred over a
period corresponding to about half-a-decade drop in temperature, 
say from about 100 MeV to 17 MeV.  Recalling that the energy density of 
black body radiation depends on temperature as $T^4$ (the 
Stefan--Boltzmann Law), we can correct roughly for the drop in temperature 
by multiplying (\ref{Exest2}) by the factor $(T_F/T_I)^{4/3}$ which 
for $T_I \sim 100 \ {\rm MeV}, T_F \sim 17\ {\rm MeV}$ gives a 
value of $\sim (17/100)^{1.33} \sim 0.095$, hence for the corrected 
expansion factor ${\cal E}$:
\begin{equation}
{\cal E}_{\rm temp.\ corr.} =\frac{{\cal R}_F}{{\cal R}_I} \sim 1.3 
   \times 10^{10},
\label{Exest3}
\end{equation}
or that the initial size of the universe before the VTR1 expansion 
as:
\begin{equation}
{\cal R}_{I} \sim  2.87 \times 10^5\ {\rm m},
\label{calRT1}
\end{equation}
and for the ratio of size to horizon: 
\begin{equation}
\frac{{\cal R}}{{\cal H}}\Bigg\vert_{T = 17\ {\rm MeV}}
   \sim  0.19; \ \ \ {\rm temp.\ corr.},
\label{Rratioiso1}
\end{equation}
leaving the universe still lying inside the horizon.

But this is about the best that we can do at present to correct for 
the isothermal approximation made, without further knowledge of the 
physics governing the universe in the earlier epoch at temperatures 
above the VTR1 transition.  Ideally, one should have followed the 
development of the universe as it evolved in time towards the VTR1 
transition to evaluate properly the expansion factor ${\cal E}$ one 
needs, but this is not possible, at least for us, when even the 
Kaluza-Klein theory supposed to govern that evolution is not fully 
spelt out.  

Given these very approximate numbers and the fact that they are, in
the first place, meant to be just order of magnitude estimates, one 
can only say that: 
\begin{itemize}
\item {\bf [C]} The expansion due to the vacuum transition VTR1 in 
FSM is of an order of magnitude capable of taking a universe inside
the horizon at some time before, to a size of the order of a light 
year at time $T_+$ just after, which will then further evolve by 
Hubble expansion to our universe today, of which all portions would 
then have, at one stage, been causally connected.
\end{itemize}  
But even so, it would seem that one has here on offer a possible 
answer to the immediate question posed by the horizon problem.

To qualify, however, as a serious candidate solution of the problem, 
it has yet to be ensured that the above scenario is not inconsistent 
with other known properties of the universe, and this has not been 
done.  There are questions that we have heard about but cannot as 
yet answer, such as the flatness problem, the problem of large scale 
inhomogeneities and so on, which the VTR1 transition might affect.  
There are probably many other questions (for example the 
entropy paradox \cite{penrose}) that we are insufficiently 
knowledgeable even to ask, let alone answer.  The purpose here is 
therefore not to claim a new solution of the horizon problem but 
merely to make known a suggestive observation from the angle of 
particle physics in the hope that experts in cosmology and the early 
universe might wish to consider.

But if, at this low level of plausibility, the above VTR1 scenario 
is taken as a candidate solution to the horizon problem, then it 
would have in addition the following seemingly attractive features:   
\begin{itemize}
\item It is not constructed specifically to solve the horizon problem 
but a consequence of the FSM constructed for particle physics, apart 
of course from the Kaluza-Klein embedding which is what connects it 
to cosmology.   Even the values of all the parameters which entered 
in the above discussion came from fits to particle physics data and 
none has been adjusted in this paper to arrive at the stated results.
\item The VTR1, being driven in the FSM by the underlying microscopic 
particle physics, would turn itself on and off automatically as the 
universe cooled and, being in the FSM a global effect, it would do so 
simultaneuosly over all space \footnote{It would seem thus not to 
require any extraneously added strategy to achieve a ``graceful exit'', 
a term used in some inflation theories to designate the need to stop 
the expansion everywhere all at once}.
\item The present FSM fit to particle data says that when $R$ reaches 
the value 1 at VTR1, it will stay there ever afterwards, meaning that 
the expansion caused by VTR1 would be all over by $t \sim 2.5 \times 
10^{-3}$ s.  It need not therefore affect previous successes of the 
Big Bang idea, such as that on nucleosynthesis, which occurred at 
much later times.
\item The expansion caused by VTR1 being ``quasi-isothermal'' and
even during that short time when it operated it is supposed to incur 
no more than the usual cooling of the universe, no reheating would 
seem to be required.  Indeed, since the overall (multidimensional) 
volume is supposed to remain almost the same, the transition can be 
interpreted less as an expansion than as a mere change of shape. 
\end{itemize}

\section{Concluding Remark}

The observation of the preceding section on the horizon problem, though 
exciting it can be, is only speculative, depending as it does on the 
Kaluza--Klein assumption, which, as far as we know has so far no empirical 
justification, and involving also some steps which can do with much 
closer scrutiny.  Nevertheless, it serves as an indication of what 
powerful effects the VTR1 transition can have on the dynamics of the 
hidden sector.  Now in the FSM, a peculiar fact is that although the 
hidden and standard sectors have little direct interactions with one 
another, their dynamics are nevertheless intimately connected.  For 
example, even the mass and mixing patterns of quarks and leptons, we 
recall from Section 2, are consequences of framon-loop renormalization 
in the hidden sector.  Recall further that the hidden sector is in FSM 
potentially where the dark matter of the universe resides.  It looks 
very likely, therefore, that the VTR1 would play a significant role 
in determining both the intrinsic properties of the underlying theory 
and the overall structure of the universe.  Unfortunately, we have as 
yet no idea how to probe systematically these effects, or even what 
salient questions to ask, our knowledge of the hidden sector being at 
present so strictly limited.  We can only hope that, with persistence 
and help from the community, something will eventually emerge to take 
us further on our way. 
\ \\
\ \\
{\large {\bf Appendix}}
\ \\

What interests us in this appendix is the behaviour of the evolution 
equations (\ref{Rdot}) and (\ref{thetadot}) near $R = 1$ where the 
vacuum transition VTR1 is seen to occur, as noted in the text.  Our 
first concern is to check this conclusion analytically, to make sure 
that it is not just a fluke of the numerical solution.  This would 
be prudent for there are zeros at the transition point both in the 
numerators and in the denominators on the right-hand side of the 2 
equations, and the behaviour of the solution is the result of some 
delicate cancellations of these zeros against one another, which 
cancellatons can depend on the direction on the $R\theta$-plane in 
which the transition point is approached.  

Before studying these cancellations in detail, let us first make 
clear that the curve traced out by $\balpha$ on the unit sphere is 
itself perfectly regular, with no singularities anywhere, in 
particular with nothing untoward happening at the pole.  What might 
seem singular there arise only from the singularities in the polar 
coordinates themselves.  That this is the case can be seen by 
casting the equation (\ref{phi}) which specifies this curve in 
terms of the Cartesian coordinates $\balpha= (x,y,z)$, with 
$x^2 + y^2 + z^2 = 1$, thus:
\begin{equation}
yz/x = a.
\label{phiCart}
\end{equation}
This is the equation of a hyperbolic paraboloid, and it is its 
intersection with the unit sphere that gives the curve we want.   
We note that Figure 1 of \cite{tfsm} shows only part of this 
curve corresponding to different values of the parameter $a$, 
but the whole intersection is a closed curve, easily obtainable 
from that figure using the symmetries of (\ref{phiCart}).  In 
the fitted trajectory in \cite{tfsm} (with $a = -0.1$), we stay 
in the upper hemisphere.  This is clearly shown in Figure 5 
there (the ``back'' view).  Since we have chosen there 
$a$ to be negative, the curve stays in the fourth and second 
quadrant above the equatorial plane, and switches to the first 
and third quadrant below the equatorial plane. (This is the 
part not shown, not being needed there).  One 
could have chosen $a$ to be positive, and it will be the other 
way round, but this is just a matter of choice, with no real 
significance.

Zooming in to the pole ($z=1$), where $y=-0.1 x$, we note that
the curve is tangent to this line through the pole, going from 
the fourth quadrant to the second quadrant in the northern 
hemisphere. It is also tangent to the geodesic (the great circle) 
through the pole with the same slope (if we look at the tangent 
plane there).  This means the curve has zero geodesic curvature 
at the pole.  The geodesic curvature changes sign as the curve 
goes through the pole, as can be seen pictorially in the figure,
as a consequence of the symmetry under $(x,y) 
\leftrightarrow  (-x,-y)$, or reflection with respect to the pole. 
As explained in \cite{tfsm}, this change of sign of the geodesic
curvature is significant in giving in FSM the physically crucial 
and theoretically intriguing result: $m_u < m_d$, despite the 
fact that the heavier generations appear in the opposite order:
$m_t \gg m_b, m_c \gg m_s$.

We have gone through the properties of the curve traced out by 
$\balpha$ in detail to highlight the fact that it is perfectly 
regular, with no singularity anywhere, so that the noted VTR1 
transition at the pole must have originated from the equations 
(\ref{Rdot}) and (\ref{thetadot}) governing the speed at which 
$\balpha$ moves along that curve, and it is to these equations 
we now turn our attention.

We shall concentrate on the region around the pole, and expand 
around that point thus: $R=1-\delta_1,\ \theta=\delta_2$, with 
both deltas small and positive, meaning that we are approaching 
that point from above in $\mu$.  And we take $\delta_1$ and 
$\delta_2$ to be independent so as to study how the behaviour 
may change depending on the direction in which that point is
approached on the $R\theta$-plane.  This behaviour is found to 
depend crucially on the denominator $D$ (\ref{D}) appearing in 
the equations: 
\begin{eqnarray}
D &=& R(1+2R) - 3R\cos^2 \theta + k(1-R)(1+2R)\nonumber \\ 
  &=& 2 \delta_1 + 2\delta_1^2 + 3 \delta_2^2 +3k \delta_1-2k 
      \delta_1^2 +O(\delta^3)
\label{Dexp}
\end{eqnarray}

Expanding similarly the other quantities appearing in the 
equations, and ignoring some unimportant positive constant 
coefficents, we find that:
\begin{eqnarray}
\dot{R} & \sim & -\frac{\delta_1 (1-\delta_1)(3-2\delta_1)}
                 {D} 
\label{Rthetadotexp1} \\ 
\dot{\theta} & \sim & -\frac{2(1-\delta_2)\delta_2}{D}
\label{Rthetadotexp2}
\end{eqnarray}

The following conclusion then results:

\begin{itemize}

\item {\bf (a)} If $\delta_1 \ll \delta_2^2$, 
then $\dot{R} \to 0, \dot{\theta} \to - \infty$.

\item {\bf (b)} If $\delta_1 \sim \delta_2^2$,  
then $\dot{R} \to$ nonzero constant, $\dot{\theta} \to \infty$.

\item {\bf (c)} If $\delta_1 \gg \delta_2^2$, 
then $\dot{R} \to$ nonzero constant, $\dot{\theta} \to 0$.

\end{itemize}

As expected, the behaviour of the trajectories for $R$ and 
$\theta$ at the transitional point depends crucially on the 
direction in which that point is approached on the 
$R\theta$-plane.  This direction, however, is not free for us 
to choose but is determined by the initial condition in solving 
the equations, or in the case of \cite{tfsm} by the fit to the 
higher mass states above the transitional point.  And this fit 
has chosen the case ${\bf (a)}$, as seen in Figure \ref{thetafit} 
by the steep approach of $\theta$ to the transition point as 
compared with the slow approach of $R$ seen in Figure \ref{Rfit}.  
This means that the condition set out in {\bf (a)} is satisfied 
and that $\dot{R} \rightarrow 0$ as the result, confirming thus 
the conclusion on VTR1 drawn from the numerical solution in 
\cite{tfsm}.  Indeeed, it is fortunate for the FSM that it has 
made this choice {\bf (a)}, for the other two cases {\bf (b)} 
and {\bf (c)} are not physical in the model where $R \leq 1$ is 
required.  

After going over the pole, the case {\bf (a)} now being chosen, 
$\dot{R} = 0$ and hence $R = 1$ for the rest of the trajectory.
This means that the tenets of the VTR1 transition first deduced 
from the numerical fit in \cite{tfsm} is analytically confirmed.

Indeed, that $\dot{R} = 0, R = 1$ allows also (\ref{thetadot}) 
to be analytically integrated and again confirms the numerical 
result obtained in \cite{tfsm}, but as this concerns only such 
results as $m_u < m_d$ already mentioned, of interest elsewhere 
but not directly in the present VTR1 context, it need not be 
detailed here.

Next, knowing now that the VTR1 transition does indeed occur in 
the scenario described in \cite{tfsm}, we ask whether this 
assertion is limited only to the 1-loop approximation studied 
there, or it is likely to persist in a more general situation.  
We intend to show in what follows that the latter is the case.

We note first that what the deduction of VTR1 in \cite{tfsm} 
relies on is the renormalization by a framon loop of a fermion 
mass matrix (not of the standard quarks and leptons but of some 
fermions in the hidden sector generically called $F$, as already
mentioned in the text), but the evolution equation (RGE) derived 
on which the result is based concerns only a left-handed vector 
pertaining to that mass matrix, given in \cite{tfsm} equation (32) 
as:
\begin{equation}
\frac{d}{dt}(\rho_S \zeta_S V_0 \bv) =
\left( \begin{array}{ccc} \gamma & 0 & 0 \\
                           0 & \gamma & 0 \\
                           0 & 0 & \beta \end{array} \right)
                          (\rho_S \zeta_S V_0 \bv).
\label{rge}
\end{equation}
That an RGE for a matrix should end up only as an evolution 
equation for a vector comes about because the fermion mass 
matrix and the framon couplings renormalizing it in this model 
are all factorizable into a left-handed (ket) vector times a 
right-handed (bra) vector.  And by the same token, it would 
remain the same no matter how many framon loops are inserted; 
the resulting RGE would be factorizable, resulting thus in 
just an evolution equation for a left-handed ket, as in 
(\ref{rge}).

Further, one notes that the evolution operator on the right of
(\ref{rge}) is a diagonal matrix with its first two entries 
indentical.  This comes about because the FSM has an inherent 
$su(2)$ symmetry between two of its framons, labelled here 
$\tilde{1}$ and $\tilde{2}$, although in \cite{tfsm} this was 
deduced by explicit calculation to 1-framon-loop order.  When 
more framon loops are inserted, the said $su(2)$ symmetry is 
maintained.  The evolution operator will have to be invariant 
under all $su(2)$ transformations of $\tilde{1}, \tilde{2}$ so 
that, by Schur's lemma, it would still have to be of the form 
shown, namely a diagonal matrix with two identical eigenvalues, 
although their values will no longer be the same as that given 
in \cite{tfsm} but would involve higher order terms.

Now in \cite{tfsm}, from the fact that two eigenvalues of the 
evolution operator are identical follows a chain of deduction 
leading to the shape equation of the curve $\Gamma$.  Thus, 
from (\ref{rge}) above, one has:
\begin{equation}
\frac{\dot{v_2}}{v_2} = \frac{\dot{v_1}}{v_1},
\label{v2dot}
\end{equation}
which gives:
\begin{equation}
v_2 = {\rm constant} \times v_1,
\label{v2}
\end{equation}
and hence, via equation (39) there,
\begin{equation}
\frac{\alpha^{\tilde{2}} \alpha^{\tilde{2}}}{\alpha^{\tilde{1}}} 
   = {\rm constant},
\label{Gamma2}
\end{equation}
This chain of deduction, relying only on the symmetry between 
$\tilde{1}$ and $\tilde{2}$ would still be valid even when these 
eigenvalues are modified by higher loops.  This means that 
$\Gamma$ would remain the same in shape, having for example a 
change in sign of its geodesic curvature at $\theta = 0$, and 
retaining thereby the result $m_u < m_d$ which we so coveted.

More critical for the occurence of the transition VTR1, however, 
are the equations (\ref{Rdot}) and (\ref{thetadot}).  Examine 
first (\ref{Rdot}).  The factor on the R.H.S. in front of the 
square bracket comes from the L.H.S. of the equation (\ref{rge}) 
and does not depend on the 1-loop calculation which affects only 
the R.H.S. of (\ref{rge}).  Now, however, the analysis above in 
(\ref{Dexp}) to (\ref{Rthetadotexp2}) of how $R$ behaves near the 
transition point relies only on this factor, not on the quantity 
inside the square bracket.  Hence, even if we were to add higher 
loop terms inside the square bracket, the behaviour of $R$ would 
remain the same and lead to the conclusions {\bf (a)} to 
{\bf (c)} as listed before, in particular {\bf (a)} under the
same conditions.  Namely, provided that the transition point is 
approached in such a way that $\delta_1 \ll \delta_2^2$, then 
$\dot{R} \to 0, \dot{\theta} \to - \infty$. 

Next, examine equation (\ref{thetadot}), or rather the equation (43) 
of \cite{tfsm} from which it is derived, namely:
\begin{equation}
\frac{\dot{\alpha}^{\tilde{3}}}{\alpha^{\tilde{3}}} 
   = \beta - \frac{\dot{P}}{P} + \frac{k}{2} \frac{\dot{R}}{R}.
\label{thetadot1}
\end{equation}
In terms of polar co-ordinates, this read as:
\begin{equation}
\dot{\theta} = - \frac{\cos \theta}{\sin \theta}  \left[ \beta 
   + \dot{R} \left(- \frac{1}{1 + 2R} + \frac{k}{2 R} 
   \right) \right].
\label{thetadot2}
\end{equation}
In these equations, all originating from (\ref{rge}), only 
$\beta$ comes from the R.H.S. of (\ref{rge}) which in the 
one-loop approximation is given in \cite{tfsm} but is the 
only quantity which will change when more loops are included.  
When the transition point is approached, as in case {\bf (a)} 
above, $\dot{R}$ vanishes, so that $\dot{\theta} \rightarrow 
- \infty$ so long as $\beta$ is finite, which it is in the 
one-loop approximation and is unlikely to change when more 
loops are added.  

We conclude therefore that the transition VTR1 is not limited 
to the one-loop appoximation but will persist when more loops 
are included.  The actual formulae will differ from \cite{tfsm} 
and so will the details of the fit to data, such as the value 
of $\mu_T$ where the transition occurs.  However, whatever new 
fit will have to respect the data and end up not very different 
from the picture shown.  We suspect, therefore, that even the 
numerical result will not change much when more loops are 
included.

\end{document}